\documentclass[journal]{IEEEtran}
\IEEEoverridecommandlockouts
\usepackage{amsmath,amssymb,amsfonts}
\usepackage{mathtools} 
\usepackage{algorithm}
\usepackage{algorithmic}
\usepackage{textcomp}
\usepackage{soul} 
\usepackage{xcolor}
\usepackage{algorithmic}
\usepackage{array}
\usepackage{subcaption}
\usepackage{stfloats}
\usepackage{url}
\usepackage{float}
\usepackage{graphicx}
\usepackage{verbatim}
\usepackage{cite} 
\usepackage{booktabs}
\usepackage[numbers,sort&compress]{natbib}
\usepackage{enumerate}
\usepackage{boondox-cal}
\usepackage{bm}  
\usepackage{CJKutf8}
\usepackage{tabularx}
\allowdisplaybreaks

\def\BibTeX{{\rm B\kern-.05em{\sc i\kern-.025em b}\kern-.08em
    T\kern-.1667em\lower.7ex\hbox{E}\kern-.125emX}}
\begin{document}

\title{GPU-Accelerated Robust Beamforming for OTFS Massive MIMO with CSI Uncertainty}

\author{Haonan Zhang, Weihua Wu
\thanks{This work was supported in part by the Natural Science Basis Research Plan in Shaanxi Province of China (Grant No. 2025JC-YBMS-674), in part by the Technological Innovation Guidance Program (Fund) - Three Reforms - Performance-Based Evaluation (Grant No. 2025ZC-YYDP-51), in part by the Shaanxi Key Industrial Innovation China Project in Industrial Domain (Grants No. 2023-ZDLGY-15, No. 2023-ZDLGY-51), in part by ``Scientist + Engineer" team building in Qinchuangyuan (Grant No. 2025QCY-KXJ-169) and in part by the Key Research and Development Program of Shaanxi (Program No.2025CY-YBXM-059).}
\thanks{Haonan Zhang, Weihua Wu, and Weijia Han are with the School of Physics and Information Technology, Shaanxi Normal University, Xi'an 710119, China (emails: zhanghaonan@snnu.edu.cn, whwu@snnu.edu.cn, wjhan@snnu.edu.cn).}
}

\maketitle
\begin{abstract}
This paper focuses on robust beamforming for orthogonal time frequency space (OTFS)-enabled massive multiple-input multiple-output (MIMO) systems under channel state information (CSI) uncertainty. In high-mobility satellite communications, uncertain CSI severely degrades beamforming accuracy and poses a major challenge to meeting users' quality of service (QoS) requirements. To address this challenge, we first formulate a chance-constrained optimization problem aiming to minimize the total transmit power while guaranteeing a predefined outage probability. Building on a 3D delay-Doppler-angle (DDA) channel representation, we propose a data-driven approach using support vector clustering (SVC) to model the uncertain CSI as an asymmetric uncertainty set. We then derive a robust counterpart that reformulates the intractable chance constraints into a deterministic semidefinite program. Finally, we design a graphics processing unit (GPU)-accelerated parallelizable alternating direction method of multipliers (ADMM) algorithm to address the computational complexity of large-scale antenna arrays. Simulation results show that the proposed SVC-based design reduces transmit power compared with conventional beamforming schemes, and that the GPU-accelerated ADMM achieves a speedup. These results confirm that the proposed framework achieves improved robustness, energy efficiency, and efficient computation in dynamic massive MIMO networks.
\end{abstract}

\section{\uppercase{{\large I}ntroduction}}
In recent years, the deployment of Low Earth Orbit (LEO) satellite constellations has become a key enabler of ubiquitous connectivity and massive access in next-generation wireless networks \cite{9210567}. Proposed systems comprise thousands of high-throughput satellites \cite{9473799}, which are intended to extend connectivity to regions that terrestrial infrastructure cannot economically cover \cite{9210567}. To date, orthogonal frequency division multiplexing (OFDM) has been the dominant transmission scheme of cellular systems since the fourth generation and remains the basis of the 5G New Radio (NR) air interface, because it splits the available bandwidth into many narrowband subcarriers and inserts a cyclic prefix, thereby converting a frequency-selective channel into a set of flat subchannels that are equalized by a single complex gain each. However, unlike terrestrial links between terminals and base stations, where the terminal velocity is low, a LEO satellite moves rapidly relative to the ground, with an orbital velocity of about 7.6~km/s at an altitude of 500~km. This motion produces a Doppler spread that introduces time selectivity, while the multipath delay spread introduces frequency selectivity, so the resulting satellite-to-terminal link is doubly selective in both time and frequency. The high relative mobility further imposes a continuous-time phase rotation and a large, rapidly varying Doppler shift on the channel, which destroys the orthogonality among subcarriers and causes severe inter-carrier interference (ICI) that conventional OFDM systems cannot handle reliably \cite{8727425,8424569}.

To overcome this bottleneck, orthogonal time frequency space (OTFS) modulation has attracted considerable research interest. By multiplexing data in the delay-Doppler (DD) domain, OTFS transforms a rapidly time-varying channel into a sparse, quasi-static two-dimensional response \cite{7925924,8424569,10152009}. Each propagation path then occupies only a few DD taps, independently of the Doppler spread. When OTFS is combined with a massive multiple-input multiple-output (MIMO) array at the satellite, the effective channel becomes jointly sparse across the delay, Doppler, and angle dimensions \cite{8727425,10916590}. This three-dimensional structure allows the transmitter to focus energy on each user's dominant directions through spatial precoding, and it supports multi-user spatial multiplexing that is markedly more robust to Doppler spread than its OFDM-based counterpart \cite{9362336,10964364}.

Despite these advantages, the beamforming accuracy of OTFS-enabled massive MIMO systems relies heavily on accurate channel state information (CSI). In practical LEO satellite communications, acquiring perfect CSI at the transmitter is infeasible owing to the substantial propagation delays over the long satellite-to-terminal range and the rapid channel variation induced by the high relative mobility \cite{10066300,10750262}. Consequently, the beamformer is designed based on the estimated channel, yet it operates over the true channel, which turns the achieved signal-to-interference-plus-noise ratio (SINR) into a random quantity determined by the CSI error. This gives rise to two failure modes. When the channel is overestimated, that is, the estimated gain exceeds the true gain, the design allocates too little power, the delivered SINR falls below its threshold, and the user experiences an outage. When the channel is underestimated, the design allocates more power than necessary; since the satellite power budget is fixed and shared among users, this waste directly reduces the number of users that can be served. Robust beamforming must therefore control both risks simultaneously rather than optimize for a single nominal channel realization. To guard against these adverse effects, the statistical distribution of the CSI must be modeled explicitly and an accurate channel uncertainty set must be constructed, so that the estimation error can be bounded mathematically and the design can be made robust to the modeled uncertainty.

Given such an uncertainty set, robustness is enforced by requiring the quality of service (QoS) constraint to hold for every channel realization in the set, which is the classical robust-optimization formulation. This formulation addresses the modeling difficulty identified above, but it does not eliminate the inherent computational challenge; instead, it transfers this burden to solving the resulting deterministic optimization problem. Traditional robust beamforming designs typically rely on predefined symmetric uncertainty sets and reformulate the chance constraints into deterministic linear matrix inequality (LMI) constraints over the uncertainty set. However, as the antenna array size and the number of users scale up, solving these high-dimensional semidefinite relaxation (SDR) problems sequentially using conventional interior-point solvers on central processing unit (CPU) incurs a rapidly growing polynomial-time complexity \cite{1166614}. This severe computational bottleneck renders traditional CPU-based solvers inadequate for the strict processing latency demanded by highly dynamic LEO networks. Consequently, there is a need to design algorithms whose computation can be decomposed into a large number of mutually independent tasks that can be executed concurrently on graphics processing unit (GPU).

Motivated by the above discussion, this paper proposes a support vector clustering (SVC)-based channel uncertainty learning aided chance-constrained robust beamforming framework with GPU-accelerated computation. We consider the downlink of a LEO satellite system in which the satellite is equipped with an $M_t$-element uniform linear array (ULA) and serves $N$ single-antenna satellite terminals (STs) within its coverage footprint, with OTFS modulation adopted to multiplex data in the DD domain. To exploit the spatial degrees of freedom of the large array, the satellite serves all STs over the same DD resources, which inevitably introduces inter-user interference: only paths whose directions of departure (DoDs) are separated by more than one array resolution in the sine domain can be spatially multiplexed without mutual interference. Specifically, the beamforming vector designed for one ST still radiates toward the paths of the other STs that fall within the same angular layer, and the corresponding signals are superposed on the intended paths in the DD domain. In addition to this inter-user interference, the system is subject to CSI uncertainty, since the long round-trip propagation delay and the rapid channel variations render the CSI at the satellite outdated and inaccurate. To achieve energy-efficient and reliable transmission, this paper formulates a chance-constrained optimization problem, where the total transmit power is minimized subject to per-ST probabilistic QoS constraints. In this problem, CSI uncertainty is characterized by an asymmetric convex uncertainty set learned from historical channel samples. The main contributions of this paper are summarized as follows:
\begin{itemize}
  \item A 3D delay-Doppler-angle (DDA) channel model is established to accurately characterize the sparse multipath components of high-mobility satellite-terrestrial links. The proposed model provides an accurate three-dimensional discrete framework tailored for the subsequent robust beamforming design.
  \item A SVC-based approach is developed to construct asymmetric convex uncertainty sets from historical channel data, achieving a characterization of CSI uncertainty in the 3D DDA domain. Building on this characterization, a robust counterpart is derived to reformulate the intractable chance constraints into a deterministic semidefinite program.
  \item A GPU-accelerated alternating direction method of multipliers (ADMM) algorithm is proposed to overcome the prohibitive computational overhead associated with large-scale antenna arrays. By decomposing the high-dimensional robust beamforming problem into a set of independent and highly parallelizable subtasks, the proposed algorithm can fully exploit the computational resources of GPUs. This design achieves substantial speedups over conventional CPU-based solvers while guaranteeing reliable QoS.
\end{itemize}
The rest of this paper is organized as follows. Section II reviews related works. Section III presents the 3D OTFS massive MIMO system model and the chance-constrained robust beamforming problem formulation. Section IV proposes the robust optimization framework based on SVC for uncertainty set learning, and elaborates on the GPU-accelerated ADMM algorithm for solving the reformulated semidefinite program. Section V provides the simulation results. Finally, Section VI concludes this paper.
\section{\uppercase{{\large R}elated {\large W}orks}}
Extensive research has been devoted to beamforming design for satellite and massive MIMO communications. For example, in \cite{9674687}, the authors proposed proposed a spatial resource allocation method based on a grid-based coverage model for massive MIMO-LEO satellite systems, where beam and time-slot resources are adaptively scheduled over coverage grids according to the non-uniform spatial distribution of capacity demands. For integrated terrestrial-satellite communications and OTFS-based MIMO transmission, the authors in \cite{9485040,10964364} developed a hybrid analog-digital beamforming scheme to mitigate inter-beam interference and enhance energy efficiency. In \cite{10859263}, a low complexity DD domain precoding scheme was proposed for multi-user MIMO-OTFS systems. Recently, neural networks have also been leveraged to enhance network utility in satellite downlink transmission. For instance, \cite{11580119} employed deep reinforcement learning to address delay-Doppler resource allocation and beamforming optimization in OTFS LEO satellite transmission, while learning-based frameworks have further been developed to coordinate interference-aware channel and power allocation across multi-beam LEO satellite downlinks \cite{11148110,10418568}. However, these studies typically rely on perfect CSI, overlooking factors such as channel estimation errors and feedback delays in practical communications, which leads to insufficient robustness of the algorithms in non-ideal environments \cite{10066300}.

To counteract the adverse effects of imperfect CSI, robust optimization methodologies have been widely investigated. Traditional robust optimization approaches characterize CSI uncertainties as predefined symmetric geometric sets, such as bounded boxes, ellipsoids, or simple polyhedra \cite{8170968,7089306}. While these models are mathematically tractable and easily integrated into standard convex solvers, their structural symmetry inherently restricts their flexibility. Distinct from traditional designs, risk-aware methods \cite{10336551} tackle CSI imperfections by considering the worst-case CSI distribution to satisfy outage probability constraints. However, both the predefined symmetric constraints and the rigid worst-case formulations tend to be overly conservative, since they inherently fail to capture the asymmetric and irregular distribution patterns of practical CSI. This structural discrepancy compels the system to provision resources for extreme scenarios that rarely occur, leading to excessively conservative strategies and severe resource waste. Alternatively, stochastic optimization approaches based on Lyapunov frameworks or stochastic differential equations (SDE) have been developed to address dynamic CSI variations \cite{9452072,8861400}. However, these techniques require instantaneous or high-frequency CSI acquisition, which is practically infeasible for long-delay links. Moreover, approaches relying on statistical characteristics \cite{10409505,10918973} depend heavily on the accuracy of assumed distributions, which restricts their applicability in highly dynamic and uncertain satellite-terrestrial environments.

Besides the modeling challenges discussed above, the computational cost of robust beamforming in massive MIMO systems remains a major concern. Existing robust approaches typically solve their SDR-based reformulations with generic interior-point solvers executed sequentially on CPUs. The complexity of such methods grows rapidly with the number of antennas and users and becomes prohibitive for the large variable dimensions induced by massive antenna arrays and OTFS frames \cite{8683524,8682193}; robust counterparts under CSI uncertainty further require repeatedly solving convex subproblems, e.g., via successive convex approximation (SCA) \cite{8683524,8901184}. Motivated by this issue, parallel computing has recently been exploited to accelerate signal processing in next-generation networks. For example, model-driven deep unfolding detectors have been developed to reduce the detection complexity of massive MIMO receivers \cite{10132441,10622446} and OTFS receivers \cite{9348493}. However, GPU-accelerated solvers for the highly coupled, non-convex robust beamforming problem have received little attention. Therefore, developing a GPU-accelerated optimization framework with high computational efficiency and robustness for OTFS massive MIMO systems remains an important challenge.

\section{\uppercase{{\large S}ystem {\large M}odel} {\large A}nd {\large 3}D {\large O}tfs {\large C}hannel}\label{sc:3}
\begin{figure}[!t]
    \centering
    \includegraphics[width=0.9\columnwidth]{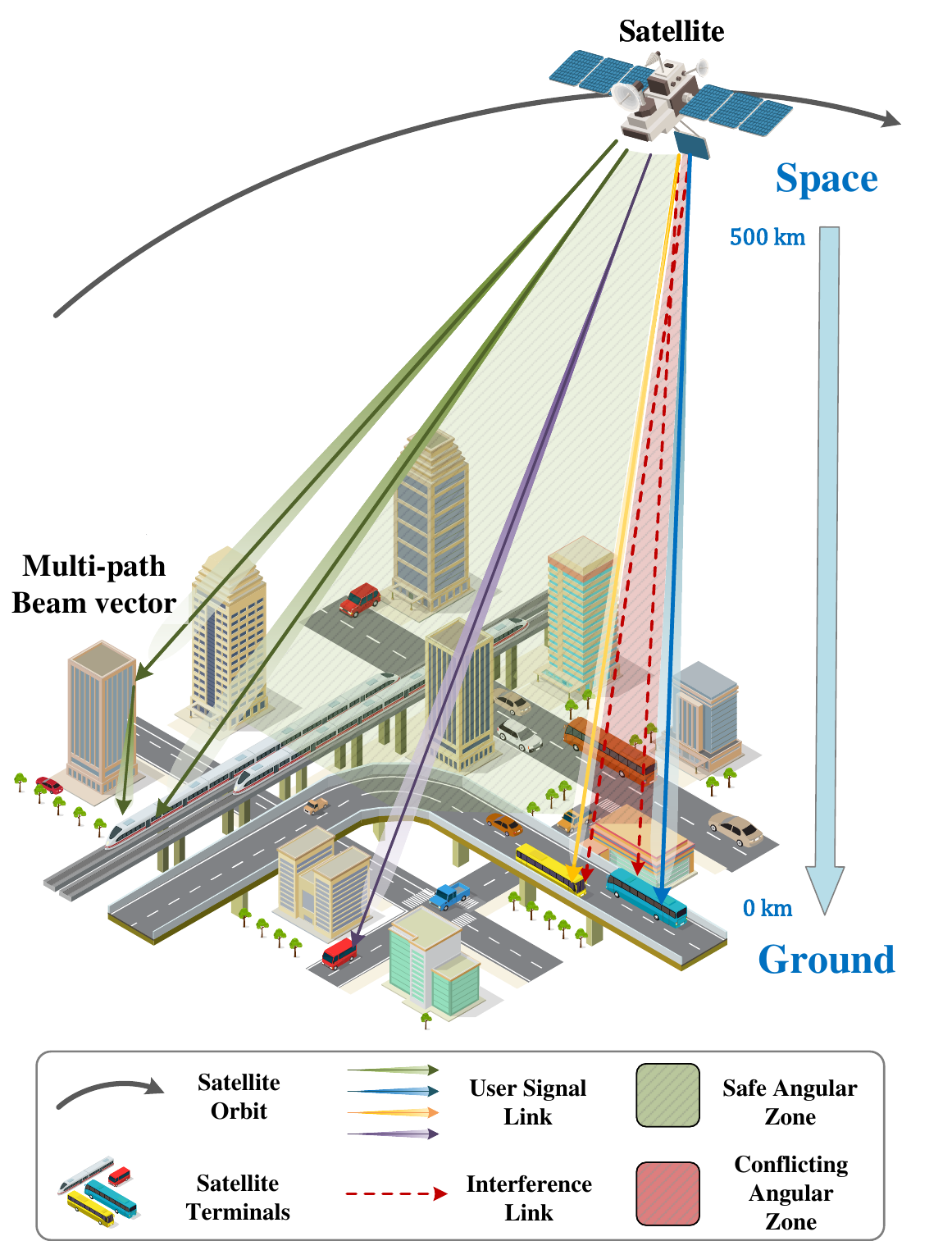}
    \caption{System model}
    \label{fig:scene}
\end{figure}

In this paper we consider a downlink massive MIMO-OTFS system for high-mobility LEO satellite communications, as illustrated in Fig. \ref{fig:scene}. The system consists of a LEO satellite equipped with an $M_t$-element ULA and $N$ single-antenna STs within its coverage footprint. With an orbital velocity of 7.6 km/s, LEO satellites induce severe time-frequency (TF) doubly-selective fading on propagation links. The inter-symbol interference (ISI) from multipath delay spread and ICI from Doppler shifts significantly degrade the reliability of conventional OFDM transmission. To address this issue, we adopt OTFS modulation, which multiplexes symbols in the DD domain and transforms fast time-varying channels into a quasi-static, sparse two-dimensional representation \cite{7925924,8424569}. By exploiting the angular-domain sparsity of large-scale arrays, we further decompose the channel across delay, Doppler and angle dimensions to form a three-dimensional DDA representation, which enables per-path spatial discrimination and provides an effective framework for robust beamforming design under channel uncertainty.
\subsection{High-Mobility MIMO Channel Model}
In high-mobility scenarios, the wireless signal transmitted from the satellite can reach a ST either via a line-of-sight (LoS) path or through reflections from scatterers such as buildings and trees. Consequently, the propagation channel between the satellite and the $n$-th ST is inherently frequency-selective and time-selective. To capture these effects, the channel is characterized by a collection of $P_n$ resolvable propagation paths.
For the $p$-th propagation path of the $n$-th ST, four physical parameters jointly characterize the path:
\begin{equation}\label{eq:path_params}
  \mathcal{P}_{n,p} = \{\tau_{n,p},\ \nu_{n,p},\ \theta_{n,p},\ \alpha_{n,p}\},
  \quad p = 1,\ldots,P_n,
\end{equation}
where $\tau_{n,p}$ and $\nu_{n,p}$ denote the propagation delay and Doppler frequency shift of the $p$-th multipath component for the $n$-th ST; $\theta_{n,p}$ represents the DoD, which is measured by the satellite ULA with respect to the array broadside; and $\alpha_{n,p}\in\mathbb{C}$ stands for the complex channel gain.

For the satellite ULA with $M_t$ elements and inter-element spacing $d = \lambda/2$, where $\lambda = c/f_c$ denotes the carrier wavelength, the array steering vector for a DoD $\theta$ is given by:
\begin{equation}\label{eq:steering}
  \mathbf{a}(\theta) = \frac{1}{\sqrt{M_t}}
  \bigl[1,\ e^{j\pi\sin\theta},\ \ldots,\ e^{j\pi(M_t-1)\sin\theta}\bigr]^T,
\end{equation}
where the normalization factor $1/\sqrt{M_t}$ ensures $\|\mathbf{a}(\theta)\|_2 = 1$. Since $\mathbf{a}(\theta)$ depends on the direction only through $\sin\theta$, the array response is a function of the sine of the DoD alone. To make this dependence explicit, we introduce the angular coordinate $\Theta\triangleq\sin\theta$. The time-varying downlink channel impulse response (CIR) from the $M_t$ satellite antennas to the $n$-th ST at continuous time $t$ is then expressed as:
\begin{equation}\label{eq:CIR}
  \mathbf{h}_n(t,\tau)=\sum_{p=1}^{P_n}\alpha_{n,p}\mathbf{a}(\theta_{n,p})
        \delta(\tau-\tau_{n,p})e^{j2\pi\nu_{n,p}t}\ \in\mathbb{C}^{M_t\times 1}.
\end{equation}
In Eq. \eqref{eq:CIR}, the term $\delta(\tau - \tau_{n,p})$ isolates the $p$-th multipath component at its specific delay, while $e^{j2\pi\nu_{n,p}t}$ captures the continuous-time phase rotation due to Doppler. For system-level processing, we discretize the CIR with sampling period $T_s = 1/B$, where $B$ is the system bandwidth. At the $q$-th sampling instant and the $\ell$-th delay tap $(\ell = 0, 1, \ldots, L-1)$, the discrete-time CIR is:
\begin{equation}\label{eq:CIR_disc}
  \mathbf{h}_n[q,\ell] = \sum_{p=1}^{P_n} \alpha_{n,p}\mathbf{a}(\theta_{n,p})
  \delta[\ell T_s-\tau_{n,p}]e^{j2\pi\nu_{n,p}qT_s},
\end{equation}
where $L = \lceil \tau_{\max} / T_s \rceil$ is the maximum number of resolvable delay taps.
In LEO satellite scenarios the orbital motion of the satellite dominates the relative movement: for a circular orbit at $h=500$\,km, $v_s=\sqrt{\mu_\oplus/(R_e+h)}\approx 7.62$\,km/s, about $137\times$ the speed of a ground user at 200 km/h, so that the satellite travels about 12.5\,m within one OTFS frame of $T_f=N_\nu/\Delta f=1.64$\,ms. The displacement is large in absolute terms, but its effect on every path parameter is negligible relative to the corresponding resolution of the system: the DoD changes by only $\Delta\theta\approx\arctan(12.5/5\times10^5)\approx1.4\times10^{-3}{}^\circ$; the Doppler drift induced by the radial acceleration $a_r\approx v_s^2/(R_e+h)\approx8.45$\,m/s$^2$ is $a_rf_cT_f/c\approx 0.09$\,Hz against a Doppler resolution $1/T_f\approx610$\,Hz; and the delay drift of the path with the largest Doppler shift is only $v_sT_fB/c=0.83$ of a delay bin. All four path parameters in Eq. \eqref{eq:path_params} are therefore constant within one OTFS frame, which justifies omitting the time index throughout this paper.

\subsection{OTFS 3D Delay-Doppler-Angle Representation}
OTFS places symbols in the DD domain, where a time varying channel becomes a sparse, quasi-static two-dimensional convolution~\cite{7925924}. Each propagation path contributes a single tap at the DD grid cell $(k_{n,p},l_{n,p})$ determined by its Doppler and delay values: $k_{n,p} = \left\lfloor \frac{\nu_{n,p}}{\Delta f_{\text{Dopp}}} \right\rfloor \in \{0, \ldots, N_\nu-1\}$ and $l_{n,p} = \left\lfloor \frac{\tau_{n,p}}{T_s} \right\rfloor \in \{0, \ldots, N_\tau-1\}$, where $N_\tau$ and $N_\nu$ are the numbers of delay and Doppler bins, respectively, and $\Delta f_{\text{Dopp}} = 1/(N_\tau N_\nu T_s)$ is the Doppler resolution.

\begin{figure}[htbp]
    \centering
    \includegraphics[width=0.9\columnwidth]{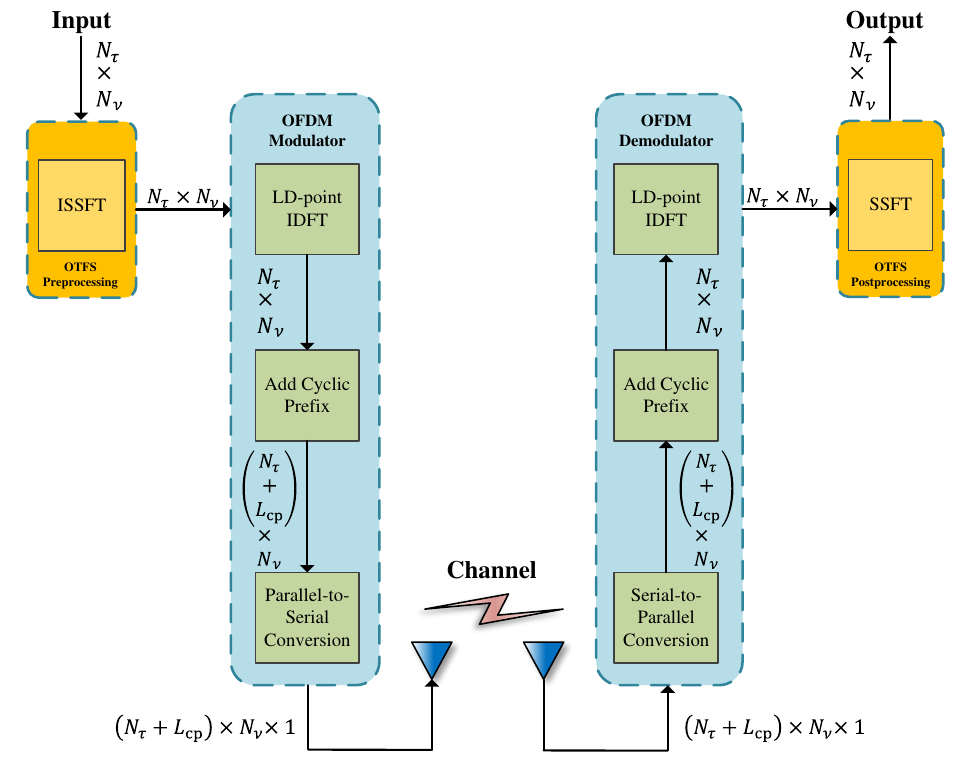}
    \caption{Block Diagram of OTFS Modulation and Demodulation}
    \label{fig:OTFS modulation}
\end{figure}

The OTFS modulation and demodulation process is depicted in Fig. \ref{fig:OTFS modulation}.
At the transmitter side, a two-dimensional block of $N_\tau \times N_\nu$ complex data symbols intended for the $n$-th ST is arranged into a matrix $\tilde{\mathbf{X}}_n^{\text{DD}} \in \mathbb{C}^{N_\tau \times N_\nu}$ in the DD domain, whose rows are indexed by the delay $l=0,\dots,N_\tau-1$ and whose columns are indexed by the Doppler $k=0,\dots,N_\nu-1$, so that the entry at row $l$ and column $k$ is $\tilde{x}_n[k,l]$ and carries one quadrature amplitude modulation (QAM) symbol; for brevity, $\tilde{x}_n[k,l]$ is referred to below as the $(k,l)$-th DD symbol of the $n$-th ST. The DD-domain symbols are transformed through the following steps. First, the DD-domain symbols are mapped to the TF domain:
\begin{equation}
\mathbf{X}_n^{\text{TF}} = \mathbf{F}_{N_\tau} \tilde{\mathbf{X}}_n^{\text{DD}} \mathbf{F}_{N_\nu}^H \in \mathbb{C}^{N_\tau \times N_\nu},
\end{equation}
where $\mathbf{F}_N$ is the normalized $N$-point discrete Fourier transform (DFT) matrix with entries $[\mathbf{F}_N]_{m,n} = \frac{1}{\sqrt{N}} e^{-j2\pi mn/N}$. Then, the TF-domain block is converted into the time-domain transmit signal. Applying an $N_\tau$-point inverse discrete Fourier transform (IDFT) to each column of $\mathbf{X}_n^{\text{TF}}$ yields
\begin{equation}
\tilde{\mathbf{S}}_n^{\text{T}} = \mathbf{F}_{N_\tau}^H\mathbf{X}_n^{\text{TF}} = \tilde{\mathbf{X}}_n^{\text{DD}}\mathbf{F}_{N_\nu}^H.
\end{equation}
After vectorization and cyclic prefix (CP) insertion, the time-domain baseband signal is $\mathbf{S}_n^{\text{T}} \in \mathbb{C}^{(N_\tau+L_{\text{cp}})N_\nu \times 1}$.
At the receiver side, after CP removal, the signal is rearranged into a 2D matrix, undergoes a Wigner transform ($N_\tau$-point DFT per column to return to the TF domain), and finally the Symplectic Finite Fourier Transform (SFFT) recovers the DD-domain received symbols:
\begin{equation}
\tilde{\mathbf{Y}}_n^{\text{DD}} = \mathbf{F}_{N_\tau}^H \mathbf{Y}_n^{\text{TF}} \mathbf{F}_{N_\nu} \in \mathbb{C}^{N_\tau \times N_\nu}
\end{equation}

\begin{figure}[htbp]
    \centering
    \includegraphics[width=0.9\columnwidth]{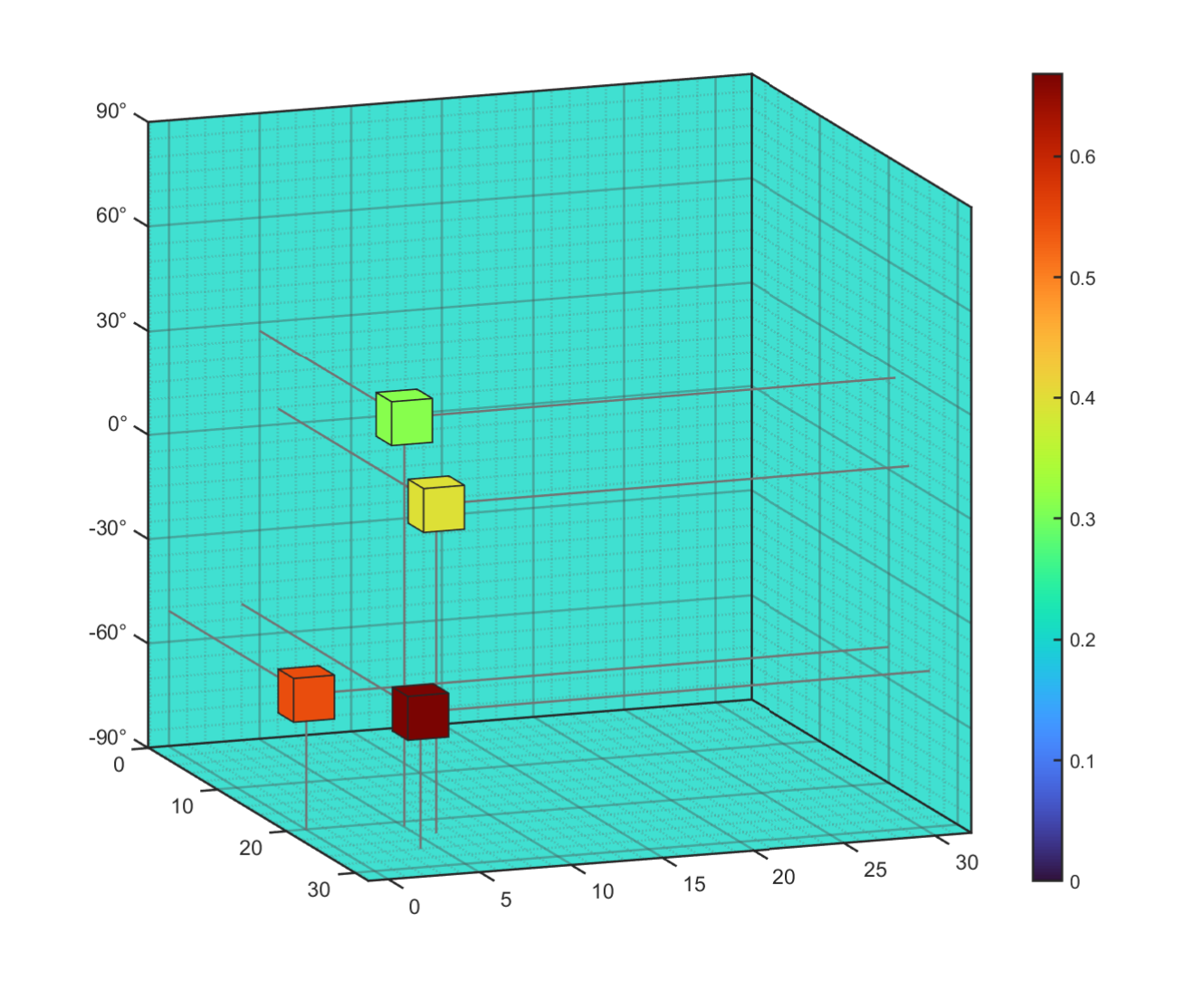}
    \caption{3D Channel}
    \label{fig:DDA}
\end{figure}

The OTFS framework described above is fundamentally a single-antenna (SISO) model. In our massive MIMO setting with an $M_t$-element ULA, the spatial dimension, parameterized by the DoD $\theta$, introduces a third axis beyond the DD plane. Consequently, the physical end-to-end channel between the satellite array and the $n$-th ST can be intuitively visualized in a three-dimensional DDA space, as illustrated in Fig. \ref{fig:DDA}.

In this 3D space, each resolvable path acts as a discrete scatterer located at the coordinate $(\tau_{n,p}, \nu_{n,p}, \theta_{n,p})$. While the 3D DDA representation provides a clear geometric interpretation, executing spatial beamforming requires mapping the DoD of each path onto the discrete angular layers of the antenna domain. Therefore, we project the 3D angular information into the spatial antenna domain. We represent the DD-domain channel as a 2D grid of spatial vectors, establishing the Delay-Doppler-Antenna (DDAnt) formulation. Specifically, the $M_t \times 1$ spatial channel vector at the discrete DD bin $[k,l]$ is formulated as:
\begin{equation}\label{eq:DDA}
  \mathbf{H}_n^{\mathrm{DDA}}[k,l] = \sum_{p=1}^{P_n} \mathbf{g}_{n,p} \delta[k-k_{n,p}]\delta[l-l_{n,p}],
\end{equation}
where $\delta[\cdot]$ denotes the discrete Kronecker delta function. The vector $\mathbf{g}_{n,p}$ encapsulates the purely spatial signature of the $p$-th path, defined by its complex gain and the ULA steering vector at its specific DoD:
\begin{equation}\label{eq:g_np}
  \mathbf{g}_{n,p} = \alpha_{n,p}\mathbf{a}(\theta_{n,p}) \in \mathbb{C}^{M_t\times 1}.
\end{equation}
This vector-valued formulation separates the delay-Doppler dispersion of the channel from its spatial characteristics. The DD coordinates of a tap determine where the contribution of the corresponding path appears on the grid, whereas its spatial signature $\mathbf{g}_{n,p}$ determines how that contribution is shaped by the transmit array. Since the data frames of all STs occupy the same DD grid, these DD coordinates carry no information that could separate one ST from another, so the beamforming design must rely on the spatial signature $\mathbf{g}_{n,p}$.
The layered structure of the three-dimensional DDA space can be intuitively interpreted as a stack of DD planes, with each angular layer corresponding to a single resolvable direction. Since the ULA response depends on the DoD only through its sine, angular separations must be measured in the coordinate $\Theta\triangleq\sin\theta$ rather than in $\theta$ itself. For an $M_t$-element ULA with spacing $d$, the first nulls of the array factor lie at $\Theta=\pm\lambda/(M_t d)$; consequently, two paths whose steering directions are separated by at least $\Delta\Theta_{\min} = \frac{2\lambda}{M_t d}$ in $\Theta$ yield approximately orthogonal array responses. They fall into distinct angular layers and remain mutually orthogonal regardless of their coordinates over the DD grid. Conversely, paths whose DoDs differ by less than $\Delta\Theta_{\min}$ in $\Theta$ occupy the same angular layer, share a strongly correlated array response, and cannot be separated spatially. This core property of the 3D DDA representation demonstrates that angular resolvability enables interference-free spatial multiplexing under ideal array conditions. We refer to this property as the path separation principle.

However, to exploit this property, the transmitter needs to acquire the DoD $\theta_{n,p}$ and the complex gain $\alpha_{n,p}$ of every path, i.e., the per-path CSI, with sufficient accuracy. In satellite communication systems, the extremely long propagation distance and high-speed relative motion between satellites and STs cause substantial round-trip time (RTT) delay and severe Doppler frequency shifts, which impair the accuracy of CSI acquisition. For this reason, perfectly accurate CSI cannot be obtained in practical scenarios, and channel estimation errors are inevitably present. Specifically, based on the physical satellite channel model established in Eq. \eqref{eq:g_np}, we define $\mathbf{g}_{n,p}$ as the true channel vector of the $p$-th multipath component corresponding to the $n$-th ST. To quantitatively characterize the imperfect CSI in practical satellite transmission, we model the channel inaccuracy via a standard additive channel error model, as presented in the following.
\begin{equation}
\mathbf{g}_{n,p}=\tilde{\mathbf{g}}_{n,p}+\mathbf{e}_{n,p},
\end{equation}
where $\tilde{\mathbf{g}}_{n,p}$ is the estimated channel vector, and $\mathbf{e}_{n,p}$ captures the estimation error.
\subsection{Signal Model and Per-Path SINR}
Having established the channel and OTFS modulation framework in Sections \ref{sc:3}-A and \ref{sc:3}-B, we now formulate the downlink signal model with spatial beamforming.

For each ST $n$, the OTFS modulator generates a time-domain baseband signal $\mathbf{s}_n^{\mathrm{T}}$ as described in Section \ref{sc:3}-B. Prior to transmission, this signal is spatially precoded by a beamforming vector $\mathbf{v}_n \in \mathbb{C}^{M_t \times 1}$. The aggregate signal vector transmitted from the $M_t$ satellite antennas at time instant $q$ is:
\begin{equation}
  \mathbf{x}[q] = \sum_{n=1}^{N}\mathbf{v}_n s_n^{\mathrm{T}}[q]\in\mathbb{C}^{M_t\times 1},
\end{equation}
where $s_n^{\mathrm{T}}[q]$ denotes the $q$-th sample of the time-domain OTFS signal and $q = 1, \ldots, N_\tau N_\nu$ indexes the effective samples within one OTFS frame. Assuming unit average symbol energy $\mathbb{E}[|s_n^{\mathrm{T}}[q]|^2] = 1$, the long-term average transmit power is $P_{\mathrm{total}} = \sum_{n=1}^{N} \mathbb{E}\bigl[|s_n^{\mathrm{T}}[q]|^2\bigr] \|\mathbf{v}_n\|^2 = \sum_{n=1}^{N} \|\mathbf{v}_n\|^2$.

After propagating through the time-varying multipath channel $\mathbf{h}_n[q,\ell]$ defined in Eq. \eqref{eq:CIR_disc}, undergoing CP removal, and passing through the OTFS demodulator (Wigner transform followed by SFFT), the received DD-domain symbol at grid cell $(k,l)$ attributed to the $p$-th path of the $n$-th ST can be expressed as:
\begin{align}\label{eq:rx}
y_{n,p}[k,l] &= \mathbf{g}_{n,p}^H\mathbf{v}_n\tilde{x}_n[k-k_{n,p},\,l-l_{n,p}]+ z_{n,p}[k,l]
\\ \nonumber &+\!\!\!\!\!\!\! \sum_{m\in\mathcal{I}(n,p)}\!\!\!\!\!\!\!\mathbf{g}_{n,p}^H\mathbf{v}_m\tilde{x}_m[k-k_{n,p},\,l-l_{n,p}],
\end{align}
where $\mathbf{g}_{n,p} = \alpha_{n,p}\mathbf{a}(\theta_{n,p})$ is the spatial channel vector of path $(n,p)$ defined in Eq. \eqref{eq:g_np}; $\tilde{x}_n[k,l]$ is the transmitted DD-domain data symbol of the $n$-th ST with $\mathbb{E}[|\tilde{x}_n[k,l]|^2] = 1$; $\mathcal{I}(n,p)$ is the set of STs whose signals are superposed on the same DD cells as path $(n,p)$ and therefore interfere with it, which is characterized in Eq. \eqref{eq:interf} below; and $z_{n,p}[k,l] \sim CN(0, \sigma^2)$ is the effective additive noise, with its variance $\sigma^2$ accounting for both the additive white Gaussian noise (AWGN) and the uncompensated fractional Doppler/delay interference in practical LEO scenarios \cite{9978028}. The integer shifts $(k_{n,p},l_{n,p})$ arise from the delay-Doppler impulse response of path $(n,p)$ and therefore apply to both the desired and the interfering data symbols, whereas the effective noise $z_{n,p}[k,l]$ is added at the receiver after demodulation and is free from any channel-induced shift.

The interference set $\mathcal{I}(n,p)$ follows directly from the path separation principle established in Section \ref{sc:3}-B. Since all STs are served over the same DD grid with full delay-Doppler reuse, the DD coordinates of a path cannot be exploited for user separation. Separation is therefore possible only in the angular dimension, and the interference set is defined from the user perspective as:
\begin{equation}\label{eq:interf}
  \mathcal{I}(n,p) \!=\! \bigl\{\! m \!\neq\! n \!:\! \exists q \!\in \!\{1,\ldots,P_m\}, |\Theta_{m,q} \!- \Theta_{n,p}| \!\leq\! \Delta\Theta_{\min} \!\bigr\}.
\end{equation}

Let $\mathbf{V}_n = \mathbf{v}_n \mathbf{v}_n^H \in \mathbb{C}^{M_t \times M_t}$ denote the rank-one beamforming matrix of the $n$-th ST. Then the SINR for the $p$-th path of the $n$-th ST is:
\begin{equation}\label{eq:SINR}
  \gamma_{n,p} = \frac{\mathbf{g}_{n,p}^H \mathbf{V}_n \mathbf{g}_{n,p}}{ \displaystyle\sum_{m \in \mathcal{I}(n,p)} \mathbf{g}_{n,p}^H \mathbf{V}_m \mathbf{g}_{n,p} + \sigma^2 } 
\end{equation} 
\subsection{Robust Beamforming Problem Formulation}
This work focuses on the optimization of beamforming vectors serving the STs. We assume that STs impose distinct QoS requirements subject to individual minimum SINR constraints. Furthermore, considering the inevitability of CSI uncertainty, we assume that these terminals can tolerate a specific maximum outage probability. Based on these premises, the robust beamforming optimization problem is formulated as follows:
\begin{subequations}\label{eq:P0}
  \begin{align}
    \!\!\!\!\!\min_{\{\mathbf{V}_n\}\succeq0} & \sum_{n=1}^N\mathrm{Tr}(\mathbf{V}_n) \label{eq:P0-a}\\
    \mathrm{s.t.}\quad&\!\!\!\!\Pr\!\left\{\gamma_{n,1} \geq \Gamma_{n,1}, \ldots, \gamma_{n,P_n} \geq \Gamma_{n,P_n}\right\}\! \geq \!1 - \epsilon, \forall n, \label{eq:P0-b} \\
    &\mathrm{rank}(\mathbf{V}_n)=1, \forall n, \label{eq:P0-c}
  \end{align}
\end{subequations}
where $\Gamma_{n,p}$ represents the QoS threshold of the $p$-th multipath of the $n$-th ST. $\epsilon$ denotes the maximum outage probability

Directly solving problem (\ref{eq:P0}) remains challenging owing to its chance constraints. To address this difficulty, we reformulate the optimization problem using uncertainty sets $\mathcal{U}_{n,p}$. Each $\mathcal{U}_{n,p}$ is built with confidence $1-\epsilon/P_n$ per path, so by the union bound over the $P_n$ paths the per-user outage probability is no larger than $\epsilon$. If the constraint $\gamma_{n,p} - \Gamma_{n,p} \geq 0$ holds for every element inside $\mathcal{U}_{n,p}$, the original chance constraints are automatically satisfied. Therefore, defining the geometry of $\mathcal{U}_{n,p}$ becomes a key step. Driven by the above considerations, we rewrite Eq. (\ref{eq:P0-b}) as follows:
\begin{equation}
    \label{eq:transformed P0-b formulation}
        \gamma_{n,p} - \Gamma_{n,p} \geq 0, \forall \mathbf{g}_{n,p}\in \mathcal{U}_{n,p},\forall n, \forall p,
\end{equation}

The SINR expression in Eq. (\ref{eq:transformed P0-b formulation}) involves quadratic forms of $\mathbf{g}_{n,p}$. Cross-multiplying and rearranging, the per-path QoS constraint $\gamma_{n,p} - \Gamma_{n,p} \geq 0$ is equivalent to:
\begin{equation}\label{eq:reformulated SINR}
  \mathbf{g}_{n,p}^H \mathbf{A}_{n,p}  \mathbf{g}_{n,p} \geq \Gamma_{n,p} \sigma^2, 
\end{equation}
where $\mathbf{A}_{n,p} = \mathbf{V}_n - \Gamma_{n,p}\sum_{m \in \mathcal{I}(n,p)} \mathbf{V}_m$. 

To address the quadratic nature of $\mathbf{g}_{n,p}$ in Eq. \eqref{eq:reformulated SINR}, we perform a Taylor expansion of $\mathbf{g}_{n,p}$ at the statistical mean $\bar{\mathbf{g}}_{n,p}$. This allows us to express the quadratic term as a linear approximation plus a remainder term:
\begin{equation}\label{eq:decomp}
  \mathbf{g}_{n,p}^H\mathbf{A}_{n,p}\mathbf{g}_{n,p} \!=\! \underbrace{2\Re\!\left\{\bar{\mathbf{g}}_{n,p}^H\mathbf{A}_{n,p}\bigl(\mathbf{g}_{n,p}\!-\!\tfrac{1}{2}\bar{\mathbf{g}}_{n,p}\bigr)\!\right\}}_{\mathrm{Linear}_{\mathbf{A}_{n,p}}(\mathbf{g}_{n,p})}\!+\!\underbrace{\bm{\delta}_{n,p}^H\mathbf{A}_{n,p}\bm{\delta}_{n,p}}_{\Delta_{n,p}},
\end{equation}
where $\mathrm{Linear}_{\mathbf{A}_{n,p}}(\mathbf{g}_{n,p})$ is the first-order linear approximation, and $\Delta_{n,p}$ is the exact remainder, with $\bm{\delta}_{n,p} = \mathbf{g}_{n,p} - \bar{\mathbf{g}}_{n,p}$.

Although the first term successfully linearizes the constraint with respect to $\mathbf{A}_{n,p}$ (and thus all beamforming matrices $\mathbf{V}_n$), simply omitting the remainder $\Delta_{n,p}$ is problematic. Because the effective matrix $\mathbf{A}_{n,p}$ is indefinite, the remainder $\Delta_{n,p}$ can be either positive or negative. If we only enforce the linearized constraint by ignoring this remainder, the system might accept a solution as feasible when the true SINR falls strictly below the required threshold $\Gamma_{n,p}$.

To ensure satisfaction of the constraint and guarantee system robustness, we derive a lower bound for the remainder and incorporate it back into the constraint. For $\Delta_{n,p}$, we have
\begin{equation}
\Delta_{n,p}=\bm\delta_{n,p}^{H}\mathbf V_n\bm\delta_{n,p}-\Gamma_{n,p}\!\!\!\!\!\!\sum_{m\in\mathcal I(n,p)}\!\!\!\!\!\!\bm\delta_{n,p}^{H}\mathbf V_m\bm\delta_{n,p}.
\end{equation}
The first term is nonnegative and can only help the constraint, so it is safe to drop for a conservative bound. All of the risk carried by the remainder comes from the interference sum, and each of its terms is bounded in the same elementary way, using $\mathbf V_m\succeq\mathbf 0$ (hence $\bm\delta^H\mathbf V_m\bm\delta\leq\lambda_{\max}(\mathbf V_m)|\bm\delta|^2$) and $\lambda_{\max}(\mathbf V_m)\leq\mathrm{Tr}(\mathbf V_m)$:
\begin{equation}\label{eq:rem_bound}
\Delta_{n,p}\geq-\Gamma_{n,p}\!\!\!\!\!\!\sum_{m\in\mathcal I(n,p)}\!\!\!\!\!\!\bm\delta_{n,p}^{H}\mathbf V_m\bm\delta_{n,p} \geq-\Gamma_{n,p}\rho_{n,p}^{2}\!\!\!\!\!\!\sum_{m\in\mathcal I(n,p)}\!\!\!\!\!\!\mathrm{Tr}(\mathbf V_m),
\end{equation}
where $\rho_{n,p}$ bounds the channel error, $\|\bm\delta_{n,p}\|_2\leq\rho_{n,p}$, and its specific evaluation is given in Section IV.

By combining (\ref{eq:decomp}) and (\ref{eq:rem_bound}), the per-path robust QoS constraint is tightened as:
\begin{align}\label{eq:corrected}
  &2\Re\Bigl\{\bar{\mathbf{g}}_{n,p}^H \Bigl( \mathbf{V}_n - \Gamma_{n,p} \!\!\!\!\!\!\!\sum_{m \in \mathcal{I}(n,p)}\!\!\!\!\!\! \mathbf{V}_m \Bigr) \Bigl( \mathbf{g}_{n,p} - \frac{1}{2}\bar{\mathbf{g}}_{n,p} \Bigr)\Bigl\} \nonumber
  \\ &\geq \Gamma_{n,p}\sigma^2 + \Gamma_{n,p}\rho_{n,p}^2 \!\!\!\!\!\!\!\!\sum_{m \in \mathcal{I}(n,p)}\!\!\!\!\!\!\! \mathrm{Tr}(\mathbf{V}_m).
\end{align}
Finally, by replacing the constraint Eq. \eqref{eq:P0-b} with Eq. \eqref{eq:corrected}, we reformulate the robust beamforming problem as follows:
\begin{subequations}\label{eq:reformulated p0}
  \begin{align}
    &\min_{\{\mathbf{V}_n\}\succeq0,}\sum_{n=1}^N\mathrm{Tr}(\mathbf{V}_n) \label{eq:reformulated p0-a}\\  \mathrm{s.t.} \quad &2\Re\left\{\bar{\mathbf{g}}_{n,p}^H\mathbf{A}_{n,p}\bigl(\mathbf{g}_{n,p}-\frac{1}{2}\bar{\mathbf{g}}_{n,p}\bigr)\right\}\geq\Gamma_{n,p}\sigma^2 \nonumber \\
&+\Gamma_{n,p}\rho_{n,p}^2\!\!\!\!\!\!\!\!\sum_{m\in\mathcal{I}(n,p)}\!\!\!\!\!\!\!\mathrm{Tr}(\mathbf{V}_m), \forall\mathbf{g}_{n,p}\in\mathcal{U}_{n,p},\forall n, \forall p, \label{eq:reformulated p0-b}\\
    &\mathrm{rank}(\mathbf{V}_n)=1,\forall n. \label{eq:reformulated p0-c}
  \end{align}
\end{subequations}

The resulting optimization problem \eqref{eq:reformulated p0} is formulated as a robust counterpart of the original problem \eqref{eq:P0}. Notably, this problem remains non-convex due to the rank-one constraint in \eqref{eq:reformulated p0-c}, which prevents the problem from being solved directly by standard convex solvers. Therefore, we apply semidefinite relaxation, analyze the rank properties of its solution in the following section, and leverage GPU-accelerated computing to solve the resulting program efficiently.

\section{\uppercase{{\large G}PU-{\large A}ccelerated {\large R}obust {\large B}eamforming} {\large O}ptimization}
Solving problem \eqref{eq:reformulated p0} poses two main difficulties: it contains a semi-infinite QoS constraint \eqref{eq:reformulated p0-b} over the uncertainty set $\mathcal{U}_{n,p}$ and a non-convex rank-one constraint \eqref{eq:reformulated p0-c}. To make it tractable, this section reformulates the problem into a finite-dimensional conic program and develops an efficient ADMM algorithm tailored for GPU computing. Specifically, Section IV-A applies linear-programming (LP) duality based on the SVC uncertainty set to reshape the semi-infinite constraint into finite linear constraints. Section IV-B then discusses the semidefinite relaxation scheme and provides the exactness conditions for obtaining a rank-one solution. Section IV-C formulates a standardized variable layout for the algorithm. Section IV-D derives the detailed ADMM updates.

\subsection{SVC-Based Uncertainty Set and Robust Counterpart}
In this subsection, we construct the uncertainty set to capture channel characterization using SVC. As a data-driven approach, SVC is designed for real-valued variables using weighted $\ell_1$ kernels and component-wise inequalities. However, the physical parameters of the considered system, including the beamforming matrices $\mathbf{V}_n$ and channel vectors, are inherently complex-valued. To address this mismatch, we propose a decoupled representation. Specifically, we preserve the complex Hermitian structure of the beamforming matrices and selectively map only the channel vectors into a real space of dimension $D=2M_t$. To facilitate this mapping, for a complex vector $\mathbf{a} \in \mathbb{C}^{M_t}$ and a complex matrix $\mathbf{B} \in \mathbb{C}^{M_t \times M_t}$, we introduce the following definitions:
\begin{align}
\check{\mathbf{a}} &\triangleq
\begin{bmatrix}\Re\{\mathbf{a}\}\\[2pt]\Im\{\mathbf{a}\}\end{bmatrix}\in\mathbb{R}^{D},
\qquad D\triangleq 2M_t,\label{eq:embed_vec}\\
\mathcal{T}(\mathbf{B}) &\triangleq
\begin{bmatrix}\Re\{\mathbf{B}\}&-\Im\{\mathbf{B}\}\\[2pt]\Im\{\mathbf{B}\}&\Re\{\mathbf{B}\}\end{bmatrix}\in\mathbb{R}^{D\times D}.\label{eq:embed_mat}
\end{align}
where $\check{(\cdot)}$ is the real-valued embedding and $\mathcal{T}(\cdot)$ its matrix form. Let $\mathbf{a}=\mathbf{a}_{\mathrm{Re}}+j\mathbf{a}_{\mathrm{Im}}$ and $\mathbf{B}=\mathbf{B}_{\mathrm{Re}}+j\mathbf{B}_{\mathrm{Im}}$, where the subscripts $\mathrm{Re}$ and $\mathrm{Im}$ denote real and imaginary parts, We then obtain that
\begin{equation}\label{eq:embedid}
  \!\!\!\!\Re\{\mathbf{a}^{H}\mathbf{B}\mathbf{b}\}=\check{\mathbf{a}}\!^{\mathsf{T}}\!\mathcal{T}\!(\mathbf{B})\check{\mathbf{b}},\,
  \check{(\mathbf{B}\mathbf{a})}=\mathcal{T}(\mathbf{B})\check{\mathbf{a}},\,
  \|\check{\mathbf{a}}\|_2=\|\mathbf{a}\|_2,
\end{equation}
and that $\mathcal{T}(\mathbf{B})$ is symmetric whenever $\mathbf{B}$ is Hermitian, while $\mathcal{T}(\cdot)$ is linear. Indeed, $\mathbf{B}\mathbf{a}=(\mathbf{B}_{\mathrm{Re}}\mathbf{a}_{\mathrm{Re}}-\mathbf{B}_{\mathrm{Im}}\mathbf{a}_{\mathrm{Im}})+j(\mathbf{B}_{\mathrm{Im}}\mathbf{a}_{\mathrm{Re}}+\mathbf{B}_{\mathrm{Re}}\mathbf{a}_{\mathrm{Im}})$ equals $\mathcal{T}(\mathbf{B})\check{\mathbf{a}}$ after stacking the real and imaginary parts; the first identity follows by expanding both sides, and the norm identity  follows from $\|\mathbf{a}\|_2^{2}=\|\mathbf{a}_{\mathrm{Re}}\|_2^{2}+\|\mathbf{a}_{\mathrm{Im}}\|_2^{2}$. Two consequences of \eqref{eq:embedid} are utilized in the subsequent derivation. First, since the embedding preserves norms, the error radius $\rho_{n,p}$ introduced in Section III-D has the same value in $\mathbb{C}^{M_t}$ and $\mathbb{R}^{D}$. Second, since $\mathcal{T}(\cdot)$ is linear and $\mathbf{A}_{n,p}$ is linear in $\{\mathbf{V}_n\}$, all embedded expressions remain affine in the optimization variables.

The CSI samples of path $(n,p)$ are stored as real vectors,
\begin{equation}\label{eq:samples}
  \mathcal{S}_{n,p}=\{\bm{\xi}_{1,n,p},\dots,\bm{\xi}_{S,n,p}\}\subset\mathbb{R}^{D},
  \bm{\xi}_{s,n,p}=\check{\mathbf{g}}^{(s)}_{n,p},
\end{equation}
where $\mathbf{g}^{(s)}_{n,p}$ is the $s$-th channel realization of path $(n,p)$ and $S$ is the sample size. The uncertainty set is learned in $\mathbb{R}^{D}$ and denoted as $\mathcal{U}_{n,p}\subset\mathbb{R}^{D}$.

Following \cite{11342326}, the samples are mapped by a nonlinear feature map $\phi(\cdot)$ into a high-dimensional space and enclosed by the smallest possible ball,
\begin{subequations}\label{eq:svcprimal}
\begin{align}
\min_{R,\bm{o},\{\varsigma_s\}} \quad & R^{2}+C\sum_{s=1}^{S}\varsigma_s \\
\text{s.t.}\quad & \bigl\|\phi(\bm{\xi}_{s,n,p})-\bm{o}\bigr\|^{2}\leq R^{2}+\varsigma_s, s=1,\dots,S\\
& \varsigma_s\geq 0, s=1,\dots,S,
\end{align}
\end{subequations}
where $R$ is the ball radius, $\bm{o}$ its center, $\varsigma_s$ are slack variables, and $C=P_n/(\epsilon S)$. With this choice of $C$, at the optimum of problem \eqref{eq:svcprimal} at most $\epsilon S/P_n$ samples fall outside the ball, so the learned set contains at least $(1-\epsilon/P_n)\times 100\%$ of the samples.

The Lagrange dual of problem \eqref{eq:svcprimal} is the convex quadratic program
\begin{subequations}\label{eq:svcdual}
\begin{align}
\min_{\bm{\lambda}} \quad & \sum_{s=1}^{S}\sum_{s'=1}^{S}\lambda_s\lambda_{s'}\,\mathrm{Ker}(\bm{\xi}_s,\bm{\xi}_{s'})
-\sum_{s=1}^{S}\lambda_s\,\mathrm{Ker}(\bm{\xi}_s,\bm{\xi}_s) \\
\text{s.t.}\quad & 0\leq\lambda_s\leq C, \quad \forall s, \\
& \sum_{s=1}^{S}\lambda_s=1,
\end{align}
\end{subequations}
where $\mathrm{Ker}(\cdot,\cdot)$ is the kernel function, and the center satisfies $\bm{o}=\sum_{s}\lambda_s\phi(\bm{\xi}_s)$. To obtain the set, we use the weighted $\ell_1$ kernel
\begin{equation}\label{eq:kernel}
\begin{split}
  \mathrm{Ker}(\mathbf{a},\mathbf{b})&=\sum_{d=1}^{D}\Xi_d-\bigl\|\mathbf{Q}_{n,p}(\mathbf{a}-\mathbf{b})\bigr\|_1,\\
  \Xi_d&=\max_{1\leq s\leq S}\mathbf{q}_d^{\mathsf{T}}\bm{\xi}_{s,n,p}-\min_{1\leq s\leq S}\mathbf{q}_d^{\mathsf{T}}\bm{\xi}_{s,n,p},
\end{split}
\end{equation}
where $\mathbf{Q}_{n,p}=\bm{\Sigma}_{n,p}^{-1/2}\in\mathbb{R}^{D\times D}$ is symmetric and nonsingular, $\bm{\Sigma}_{n,p}$ is the sample covariance of $\mathcal{S}_{n,p}$, and $\mathbf{q}_d$ is the $d$-th column of $\mathbf{Q}_{n,p}$.

Solving problem \eqref{eq:svcdual} gives the multipliers $\lambda_{s,n,p}$ and partitions the samples into the support vector (SV) set $\mathcal{F}_{n,p}=\{s|\lambda_{s,n,p}>0\}$ and the boundary SV set $\mathcal{B}_{n,p}=\{s|0<\lambda_{s,n,p}<C\}\subseteq\mathcal{F}_{n,p}$, whose cardinality we write as $K_{n,p}\triangleq|\mathcal{F}_{n,p}|$. The learned set is $\{\check{\mathbf{g}}:\|\phi(\check{\mathbf{g}})-\bm{o}\|^{2}\leq R^{2}\}$, i.e.
\begin{equation}\label{eq:learnedset}
\begin{split}
  &\mathrm{Ker}(\check{\mathbf{g}},\check{\mathbf{g}})-2\sum_{s\in\mathcal{F}_{n,p}}\lambda_s\,\mathrm{Ker}(\check{\mathbf{g}},\bm{\xi}_s)\\
  &+\sum_{s,s'\in\mathcal{F}_{n,p}}\lambda_s\lambda_{s'}\,\mathrm{Ker}(\bm{\xi}_s,\bm{\xi}_{s'})\leq R^{2},
\end{split}
\end{equation}
where $R^{2}=\|\phi(\bm{\xi}_s)-\bm{o}\|^{2}$ is evaluated at any boundary SV $\bm{\xi}_s$, $s\in\mathcal{B}_{n,p}$.

Substituting Eq. \eqref{eq:kernel} into Eq. \eqref{eq:learnedset} and using $\mathrm{Ker}(\mathbf{a},\mathbf{a})=\sum_{d=1}^{D}\Xi_d$ and $\sum_{s}\lambda_s=1$, the constant terms cancel out and Eq. \eqref{eq:learnedset} reduces to
\begin{equation}\label{eq:l1set}
  \sum_{s\in\mathcal{F}_{n,p}}\lambda_{s,n,p}\bigl\|\mathbf{Q}_{n,p}(\check{\mathbf{g}}-\bm{\xi}_{s,n,p})\bigr\|_1\leq\varrho_{n,p},
\end{equation}
\begin{equation}\label{eq:l1radius}
  \varrho_{n,p}\triangleq\min_{k\in\mathcal{B}_{n,p}}\sum_{s\in\mathcal{F}_{n,p}}\lambda_{s,n,p}\bigl\|\mathbf{Q}_{n,p}(\bm{\xi}_{k,n,p}-\bm{\xi}_{s,n,p})\bigr\|_1,
\end{equation}
where $\varrho_{n,p}$ is a constant to ensure that the above
inequality is always valid.

Introducing auxiliary vectors $\bm{\mu}_s\in\mathbb{R}^{D}$, $s\in\mathcal{F}_{n,p}$, that upper-bound the absolute values component-wise, the set can be written as
\begin{equation}\label{eq:lifted}
  \mathcal{U}_{n,p}\!=\!\left\{\!\check{\mathbf{g}}\in\mathbb{R}^{D}\!\left|
  \begin{aligned}
    &\mathbf{Q}_{n,p}(\check{\mathbf{g}}-\bm{\xi}_s)\leq\bm{\mu}_s,\forall s\in\mathcal{F}_{n,p}\\
    &\!-\!\mathbf{Q}_{n,p}(\check{\mathbf{g}}-\bm{\xi}_s)\leq\bm{\mu}_s,\forall s\in\mathcal{F}_{n,p}\\
    &\textstyle\sum_{s\in\mathcal{F}_{n,p}}\lambda_{s,n,p}\mathbf{1}_D^{\mathsf{T}}\bm{\mu}_s\leq\varrho_{n,p}
  \end{aligned}\!\right\},\right.
\end{equation}
where $\mathbf{1}_D$ is the all-ones vector of length $D$.

The tightening term in Eq. \eqref{eq:reformulated p0-b} uses the Euclidean radius of the same set,
\begin{equation}\label{eq:rhodef}
\rho_{n,p}\triangleq\max_{\check{\mathbf{g}}\in\mathcal{U}_{n,p}}\bigl\|\check{\mathbf{g}}-\check{\bar{\mathbf{g}}}_{n,p}\bigr\|_2,
\end{equation}
where $\check{\bar{\mathbf{g}}}_{n,p}$ is the embedded statistical mean channel. By Eq.~\eqref{eq:embedid}, $\rho_{n,p}$ is also the largest Euclidean norm of the channel error $\bm{\delta}_{n,p}=\mathbf{g}_{n,p}-\bar{\mathbf{g}}_{n,p}$ over $\mathcal{U}_{n,p}$. Computing Eq.~\eqref{eq:rhodef} exactly requires maximizing a convex function over a polytope, which is costly; however, $\rho_{n,p}$ enters Eq. \eqref{eq:reformulated p0-b} with a nonpositive coefficient, so any upper bound keeps the constraint valid and only increases the conservatism. From $\|\mathbf{Q}\mathbf{a}\|_1\geq\|\mathbf{Q}\mathbf{a}\|_2\geq\sigma_{\min}(\mathbf{Q})\|\mathbf{a}\|_2$ and the triangle inequality, Eq. \eqref{eq:l1set} implies $\sum_{s}\lambda_s\|\check{\mathbf{g}}-\bm{\xi}_s\|_2\leq\varrho_{n,p}/\sigma_{\min}(\mathbf{Q}_{n,p})$ for every $\check{\mathbf{g}}\in\mathcal{U}_{n,p}$, and hence
\begin{equation}\label{eq:rhobound}
  \rho_{n,p}\leq\hat{\rho}_{n,p}\!\triangleq\!\frac{\varrho_{n,p}}{\sigma_{\min}(\mathbf{Q}_{n,p})}+\!\!\!\!\sum_{s\in\mathcal{F}_{n,p}}\!\lambda_{s,n,p}\bigl\|\bm{\xi}_{s,n,p}-\check{\bar{\mathbf{g}}}_{n,p}\bigr\|_2,
\end{equation}
where $\sigma_{\min}(\cdot)$ denotes the smallest singular value. In the implementation we set $\rho_{n,p}=\hat{\rho}_{n,p}$.

Recall from Section III-D the definition $\mathbf{A}_{n,p}=\mathbf{V}_n-\Gamma_{n,p}\sum_{m\in\mathcal{I}(n,p)}\mathbf{V}_m$, where $\mathcal{I}(n,p)$ is the set of paths interfering with path $(n,p)$. Applying the first identity in Eq.~\eqref{eq:embedid} to the left-hand side of Eq.~\eqref{eq:reformulated p0-b} gives
\begin{equation}\label{eq:expansion}
  2\Re\Bigl\{\bar{\mathbf{g}}^{H}\mathbf{A}\bigl(\mathbf{g}-\frac{1}{2}\bar{\mathbf{g}}\bigr)\Bigr\}=2\bigl(\mathcal{T}(\mathbf{A})\check{\bar{\mathbf{g}}}\bigr)^{\mathsf{T}}\check{\mathbf{g}}
  -\check{\bar{\mathbf{g}}}^{\mathsf{T}}\mathcal{T}(\mathbf{A})\check{\bar{\mathbf{g}}},
\end{equation}
where the path indices are dropped for readability. Substituting Eq.~\eqref{eq:expansion} into Eq.~\eqref{eq:reformulated p0-b} and moving the terms that do not depend on $\check{\mathbf{g}}$ to the right-hand side, we obtain
\begin{equation}\label{eq:compact}
  \mathbf{p}_{n,p}(\mathbf{V})^{\mathsf{T}}\check{\mathbf{g}}\le x_{n,p}(\mathbf{V}),\qquad\forall\,\check{\mathbf{g}}\in\mathcal{U}_{n,p},
\end{equation}
where
\begin{align}
\mathbf{p}_{n,p}(\mathbf{V}) &=-2\mathcal{T}(\mathbf{A}_{n,p})\check{\bar{\mathbf{g}}}_{n,p}\in\mathbb{R}^{D},\label{eq:pdef}\\
x_{n,p}(\mathbf{V}) &=
-\check{\bar{\mathbf{g}}}_{n,p}^{\mathsf{T}}\mathcal{T}(\mathbf{A}_{n,p})\check{\bar{\mathbf{g}}}_{n,p}
-\Gamma_{n,p}\sigma^{2}\nonumber \\
&\quad-\Gamma_{n,p}\rho_{n,p}^{2}\sum_{m\in\mathcal{I}(n,p)}\mathrm{Tr}(\mathbf{V}_m)\in\mathbb{R}.
\label{eq:xdef}
\end{align}

Since $\mathbf{A}_{n,p}$ is linear in $\{\mathbf{V}_n\}$ and both $\mathcal{T}(\cdot)$ and $\mathrm{Tr}(\cdot)$ are linear, $\mathbf{p}_{n,p}(\mathbf{V})$ is linear and $x_{n,p}(\mathbf{V})$ is affine in the beamforming matrices, which keeps the final problem a linear semidefinite program. The last term of Eq.~\eqref{eq:xdef} is the tightening term; it only adds the coefficient $-\Gamma_{n,p}\rho_{n,p}^{2}$ to the trace of every interfering user's matrix.

Constraint Eq.~\eqref{eq:compact} holds for all $\check{\mathbf{g}}\in\mathcal{U}_{n,p}$ if and only if $\max_{\check{\mathbf{g}}\in\mathcal{U}_{n,p}}\mathbf{p}^{\mathsf{T}}\check{\mathbf{g}}\leq x$. Using Eq.~\eqref{eq:lifted}, this worst case is the linear program
\begin{subequations}\label{eq:worstlp}
\begin{align}
  &\max_{\check{\mathbf{g}},\{\bm{\mu}_s\}} \mathbf{p}^{\mathsf{T}}\check{\mathbf{g}}\\
  \text{s.t.}\quad & \mathbf{Q}(\check{\mathbf{g}}-\bm{\xi}_s)-\bm{\mu}_s\leq\mathbf{0},\\
  & -\mathbf{Q}(\check{\mathbf{g}}-\bm{\xi}_s)-\bm{\mu}_s\leq\mathbf{0},\\
  & \textstyle\sum_{s\in\mathcal{F}}\lambda_s\mathbf{1}_D^{\mathsf{T}}\bm{\mu}_s-\varrho\leq 0,
\end{align}
\end{subequations}

The Lagrangian of problem \eqref{eq:worstlp} is
\begin{equation}\label{eq:lag}
\begin{split}
  L\!&=\!\mathbf{p}\!^{\mathsf{T}}\check{\mathbf{g}}
  \!+\!\!\sum_{s\in\mathcal{F}}\bm{\omega}_s^{\mathsf{T}}\bigl(\bm{\mu}_s\!-\!\mathbf{Q}(\check{\mathbf{g}}\!-\!\bm{\xi}_s)\bigr)
  \!+\!\!\sum_{s\in\mathcal{F}}\bm{\psi}_s^{\mathsf{T}}\bigl(\bm{\mu}_s\!+\!\mathbf{Q}(\check{\mathbf{g}}\!-\!\bm{\xi}_s)\bigr)\\
  &\quad+\chi\Bigl(\varrho-\sum_{s\in\mathcal{F}}\lambda_s\mathbf{1}_D^{\mathsf{T}}\bm{\mu}_s\Bigr).
\end{split}
\end{equation}
where the nonnegative multipliers $\bm{\omega}s\in\mathbb{R}+^{D}$, $\bm{\psi}s\in\mathbb{R}+^{D}$ and $\chi\in\mathbb{R}_+$ are associated with the three families of inequalities in Eq.~\eqref{eq:worstlp}, respectively. 
Collecting the terms that multiply $\check{\mathbf{g}}$ and $\bm{\mu}_s$, and using $\mathbf{Q}^{\mathsf{T}}=\mathbf{Q}$, we obtain the dual problem
\begin{subequations}\label{eq:dual}
\begin{align}
  \min_{\chi,\{\bm{\omega}_s\},\{\bm{\psi}_s\}}\quad
  &\sum_{s\in\mathcal{F}}(\bm{\omega}_s-\bm{\psi}_s)^{\mathsf{T}}\mathbf{Q}\bm{\xi}_s+\varrho\,\chi\\
  \text{s.t.}\quad
  &\mathbf{Q}\sum_{s\in\mathcal{F}}(\bm{\omega}_s-\bm{\psi}_s)=\mathbf{p},\\
  &\bm{\omega}_s+\bm{\psi}_s=\lambda_s\chi\mathbf{1}_D,\quad\forall s\in\mathcal{F},\\
  &\chi\geq 0,\quad\bm{\omega}_s\geq\mathbf{0},\quad\bm{\psi}_s\geq\mathbf{0}.
\end{align}
\end{subequations}

Since $\mathcal{U}_{n,p}$ is compact, the LP problem \eqref{eq:worstlp} attains its optimum, and strong duality holds for linear programs. Therefore Eq.~\eqref{eq:compact} holds if and only if there exist $\chi_{n,p}\geq 0$ and $\bm{\omega}_{s,n,p},\bm{\psi}_{s,n,p}\geq\mathbf{0}$, $s\in\mathcal{F}_{n,p}$, satisfying
\begin{subequations}
\begin{align}
  &\sum_{s\in\mathcal{F}_{n,p}}\bigl(\bm{\omega}_{s,n,p}-\bm{\psi}_{s,n,p}\bigr)^{\mathsf{T}}\mathbf{Q}_{n,p}\bm{\xi}_{s,n,p}\nonumber\\
  &+\varrho_{n,p}\chi_{n,p}\leq x_{n,p}(\mathbf{V}),\label{eq:rc-a}\\
  &\mathbf{Q}_{n,p}\sum_{s\in\mathcal{F}_{n,p}}\bigl(\bm{\omega}_{s,n,p}-\bm{\psi}_{s,n,p}\bigr)=\mathbf{p}_{n,p}(\mathbf{V}),\label{eq:rc-b}\\
  &\bm{\omega}_{s,n,p}+\bm{\psi}_{s,n,p}=\lambda_{s,n,p}\,\chi_{n,p}\,\mathbf{1}_D,\quad\forall s\in\mathcal{F}_{n,p},\label{eq:rc-c}\\
  &\chi_{n,p}\geq 0,\quad\bm{\omega}_{s,n,p}\geq\mathbf{0},\quad\bm{\psi}_{s,n,p}\geq\mathbf{0},\label{eq:rc-d}
\end{align}
\end{subequations}
for every $n$ and $p$. Substituting Eq.~\eqref{eq:rc-a}--\eqref{eq:rc-d} into Eq.~\eqref{eq:reformulated p0} gives the deterministic problem solved in the remainder of this section as follows:
\begin{subequations}\label{eq:P1}
\begin{align}
  &\min_{\{\mathbf{V}_n\}\succeq\mathbf{0},\{\chi_{n,p},\bm{\omega}_{s,n,p},\bm{\psi}_{s,n,p}\}}
  \sum_{n=1}^{N}\mathrm{Tr}(\mathbf{V}_n) \label{eq:P1-a}\\
  \text{s.t.}\quad
  &\eqref{eq:rc-a}\text{--}\eqref{eq:rc-d},\forall n,\forall p, \label{eq:P1-b}\\
  &\mathrm{rank}(\mathbf{V}_n)=1,\forall n.\label{eq:P1-c}
\end{align}
\end{subequations}
Omitting the non-convex rank-one constraint Eq.~\eqref{eq:P1-c} yields a convex relaxation of problem \eqref{eq:P1}. This relaxed formulation is completely characterized by a linear objective, linear equality and inequality constraints Eq.~\eqref{eq:rc-a}-\eqref{eq:rc-d}, and a single positive semidefinite cone constraint $\mathbf{V}_n\succeq\mathbf{0}$. As a standard linear semidefinite program (SDP), this problem guarantees a globally optimal solution and forms the basis for our algorithmic design. To validate this relaxation, Section IV-B will subsequently provide the exactness conditions under which the SDP automatically yields a rank-one matrix. 

\subsection{Analysis of Solution Properties Based on Matrix Transformations}
The rank-one constraint Eq.~\eqref{eq:P1-c} is the only non-convex element in problem \eqref{eq:P1}. Consequently, this subsection analyzes the rank properties of the semidefinite relaxation formulated in the previous step.

\textbf{Assumption~1:} { \itshape The relaxed problem is strictly feasible, i.e., it has a feasible point with $\mathbf{V}_n\succ\mathbf{0}$ for every $n$. This is the standard Slater condition and holds whenever the QoS targets are jointly achievable. }

We also use $\Gamma_{n,p}>0$, $\sigma^{2}>0$, and $P_n\ge1$, which hold by the problem formulation, and $n\notin\mathcal{I}(n,p)$, which follows from Eq.~\eqref{eq:interf}. Under Assumption~1, strong duality holds and the Karush-Kuhn-Tucker (KKT) conditions are necessary and sufficient. Let $\nu_{n,p}\ge0$ be the multiplier of Eq.~\eqref{eq:rc-a}, let $\mathbf{z}_{n,p}\in\mathbb{R}^{D}$ be the multiplier of Eq.~\eqref{eq:rc-b}, and let $\mathbf{Y}_n\succeq\mathbf{0}$ be the multiplier of $\mathbf{V}_n\succeq\mathbf{0}$. The multipliers of Eq.~\eqref{eq:rc-c} and of the nonnegativities in Eq.~\eqref{eq:rc-d} are eliminated below and never appear explicitly.

Stationarity of the Lagrangian with respect to $\chi_{n,p}$, $\bm{\omega}_{s,n,p}$, and $\bm{\psi}_{s,n,p}$ reproduces the KKT conditions of the worst-case LP problem \eqref{eq:worstlp} and its dual problem \eqref{eq:dual}. Therefore, whenever $\nu_{n,p}>0$, the scaled multiplier
\begin{equation}\label{eq:gstar}
  \check{\mathbf{g}}_{n,p}^{\star}\triangleq-\frac{\mathbf{z}_{n,p}}{\nu_{n,p}}
\end{equation}
is a maximizer of $\mathbf{p}_{n,p}(\mathbf{V}^{\star})^{\mathsf{T}}\check{\mathbf{g}}$ over $\mathcal{U}_{n,p}$. In Eq.~\eqref{eq:gstar}, $\mathbf{g}^{\star}_{n,p}$ is the complex channel whose embedding is $\check{\mathbf{g}}^{\star}_{n,p}$, and its error $\bm{\delta}^{\star}_{n,p}=\mathbf{g}^{\star}_{n,p}-\bar{\mathbf{g}}_{n,p}$ satisfies $\|\bm{\delta}^{\star}_{n,p}\|_2\le\rho_{n,p}$ by Eq.~\eqref{eq:rhodef}. The linearized matrix of path $(n,p)$ is
\begin{equation}\label{eq:glin}
  \mathbf{G}^{\mathrm{lin}}_{n,p}\triangleq\mathbf{g}^{\star}_{n,p}\mathbf{g}^{\star H}_{n,p}-\bm{\delta}^{\star}_{n,p}\bm{\delta}^{\star H}_{n,p},
\end{equation}
which satisfies $\mathrm{Tr}(\mathbf{A}_{n,p}\mathbf{G}^{\mathrm{lin}}_{n,p})\!=\!2\Re\{\bar{\mathbf{g}}_{n,p}^{H}\mathbf{A}_{n,p}(\mathbf{g}^{\star}_{n,p}-\frac{1}{2}\bar{\mathbf{g}}_{n,p})\}$.

Stationarity of the Lagrangian with respect to $\mathbf{V}_n$ yields the decomposition
\begin{equation}\label{eq:yslack}
  \mathbf{Y}_n=\mathbf{B}_n-\mathbf{S}_n,
\end{equation}
\begin{subequations}\label{eq:slackparts}
\begin{align}
\mathbf{S}_n&\triangleq\sum_{p=1}^{P_n}\nu_{n,p}\,\mathbf{g}^{\star}_{n,p}\mathbf{g}^{\star H}_{n,p}\succeq\mathbf{0},\label{eq:Sndef}\\
\mathbf{B}_n&\triangleq\varphi_n\mathbf{I}\!+\!\!\!\!\!\!\sum_{(m,q)\in\mathcal{J}_n}\!\!\!\!\!\Gamma_{m,q}\nu_{m,q}\mathbf{G}^{\mathrm{lin}}_{m,q}+\sum_{p=1}^{P_n}\nu_{n,p}\bm{\delta}^{\star}_{n,p}\bm{\delta}^{\star H}_{n,p},\label{eq:Bndef}\\
\varphi_n&\triangleq1+\sum_{(m,q)\in\mathcal{J}_n}\Gamma_{m,q}\nu_{m,q}\rho_{m,q}^{2},\label{eq:phidef}
\end{align}
\end{subequations}
where $\mathcal{J}_n\triangleq\{(m,q):n\in\mathcal{I}(m,q)\}$ is the set of paths interfered by the $n$-th ST. The matrix $\mathbf{B}_n$ is always positive definite. For any unit vector $\mathbf{u}$, $\mathbf{u}^{H}\mathbf{G}^{\mathrm{lin}}_{n,p}\mathbf{u}=|\mathbf{g}^{\star H}_{n,p}\mathbf{u}|^{2}-|\bm{\delta}^{\star H}_{n,p}\mathbf{u}|^{2}\ge-\|\bm{\delta}^{\star}_{n,p}\|_2^{2}\ge-\rho_{n,p}^{2}$, so $\mathbf{G}^{\mathrm{lin}}_{n,p}\succeq-\rho_{n,p}^{2}\mathbf{I}$. Dropping the last sum in \eqref{eq:Bndef}, which is positive semidefinite, gives
\begin{equation}\label{eq:Bbound}
  \mathbf{B}_n\succeq\Bigl(\varphi_n-\sum_{(m,q)\in\mathcal{J}_n}\Gamma_{m,q}\nu_{m,q}\rho_{m,q}^{2}\Bigr)\mathbf{I}=\mathbf{I}\succ\mathbf{0},
\end{equation}
where the last equality is Eq.~\eqref{eq:phidef}. The $\rho_{m,q}^{2}$ coefficients enter $\varphi_n$ through the tightening term in Eq.~\eqref{eq:xdef}, and without them Eq.~\eqref{eq:Bbound} would hold only under an extra condition on the multipliers.

\textbf{Lemma~1:}\label{lem:nullity} { \itshape Let $\mathbf{Z}=\mathbf{A}-\mathbf{S}$ with $\mathbf{A}\succ\mathbf{0}$ and $\mathbf{S}\succeq\mathbf{0}$. If $\mathbf{Z}\succeq\mathbf{0}$, then $\dim\mathrm{null}(\mathbf{Z})\le\mathrm{rank}(\mathbf{S})$. }

\textbf{Proof:} If $\mathbf{x}\in\mathrm{null}(\mathbf{Z})\cap\mathrm{null}(\mathbf{S})$, then $\mathbf{A}\mathbf{x}=(\mathbf{Z}+\mathbf{S})\mathbf{x}=\mathbf{0}$, which forces $\mathbf{x}=\mathbf{0}$ because $\mathbf{A}\succ\mathbf{0}$. Hence $\mathbf{S}$ restricted to $\mathrm{null}(\mathbf{Z})$ is injective, and $\dim\mathrm{null}(\mathbf{Z})\le\mathrm{rank}(\mathbf{S})$ follows. $\blacksquare$

With $\mathcal{P}^{\mathrm{act}}_n\triangleq\{p:\nu_{n,p}>0\}$ the set of active paths of the $n$-th ST, the main rank bound is as follows.

\textbf{Theorem~1 (Rank bound):}\label{thm:rank} { \itshape Under Assumption~1, every optimal solution of the relaxed problem satisfies $1\le\mathrm{rank}(\mathbf{V}_n^{\star})\le\mathrm{rank}(\mathbf{S}_n)\le\min(|\mathcal{P}^{\mathrm{act}}_n|,M_t)$. }

\textbf{Proof:} Complementarity condition gives $\mathbf{Y}_n\mathbf{V}_n^{\star}=\mathbf{0}$, hence $\mathrm{range}(\mathbf{V}_n^{\star})\subseteq\mathrm{null}(\mathbf{Y}_n)$. Applying Lemma~\ref{lem:nullity} to \eqref{eq:yslack} gives $\dim\mathrm{null}(\mathbf{Y}_n)\le\mathrm{rank}(\mathbf{S}_n)$, and $\mathrm{rank}(\mathbf{S}_n)\le\min(|\mathcal{P}^{\mathrm{act}}_n|,M_t)$ because $\mathbf{S}_n$ is the sum of $|\mathcal{P}^{\mathrm{act}}_n|$ rank-one matrices. It remains to show that $\mathbf{V}_n^{\star}\neq\mathbf{0}$. Since the system \eqref{eq:rc-a}--\eqref{eq:rc-d} is equivalent to \eqref{eq:reformulated p0-b} by LP duality, writing $\mathbf{g}^{H}\mathbf{A}_{n,p}\mathbf{g}=2\Re\{\bar{\mathbf{g}}_{n,p}^{H}\mathbf{A}_{n,p}(\mathbf{g}-\frac{1}{2}\bar{\mathbf{g}}_{n,p})\}+\bm{\delta}^{H}\mathbf{A}_{n,p}\bm{\delta}$ and using $\bm{\delta}^{H}\mathbf{A}_{n,p}\bm{\delta}\ge-\Gamma_{n,p}\rho_{n,p}^{2}\sum_{m\in\mathcal{I}(n,p)}\mathrm{Tr}(\mathbf{V}_m)$ shows that the robust constraint forces $\mathbf{g}^{H}\mathbf{A}_{n,p}\mathbf{g}\ge\Gamma_{n,p}\sigma^{2}>0$ for every $\mathbf{g}\in\mathcal{U}_{n,p}$. If $\mathbf{V}_n=\mathbf{0}$, then $\mathbf{A}_{n,p}\preceq\mathbf{0}$, which is impossible. Hence $\mathrm{rank}(\mathbf{V}_n^{\star})\ge1$. $\blacksquare$

\textbf{Proposition~1 (Rank-one condition):}\label{prop:rank1} { \itshape Consider the generalized eigenproblem $\mathbf{S}_n\mathbf{u}=\theta\,\mathbf{B}_n\mathbf{u}$. Under Assumption~1, its largest generalized eigenvalue equals one. If this eigenvalue is simple, then $\mathrm{rank}(\mathbf{V}_n^{\star})=1$ and $\mathbf{V}_n^{\star}=\mathbf{v}_n^{\star}\mathbf{v}_n^{\star H}$, where $\mathbf{v}_n^{\star}$ is the principal generalized eigenvector scaled so that the active QoS constraints hold with equality. }

\textbf{Proof:} Applying a congruence transformation to \eqref{eq:yslack} with the matrix $\mathbf{B}_n^{-1/2}$ yields the inequality $\mathbf{I} - \mathbf{B}_n^{-1/2}\mathbf{S}_n\mathbf{B}_n^{-1/2}\succeq\mathbf{0}$. Hence the eigenvalues of $\mathbf{B}_n^{-1/2}\mathbf{S}_n\mathbf{B}_n^{-1/2}$, which are the generalized eigenvalues of $(\mathbf{S}_n,\mathbf{B}_n)$, are at most one, and $\dim\mathrm{null}(\mathbf{Y}_n)$ equals the multiplicity of the eigenvalue one. This eigenvalue is attained because $\mathbf{V}_n^{\star}\neq\mathbf{0}$ and $\mathrm{range}(\mathbf{V}_n^{\star})\subseteq\mathrm{null}(\mathbf{Y}_n)$ by Theorem~\ref{thm:rank}. If it is simple, then $\dim\mathrm{null}(\mathbf{Y}_n)=1$, so $\mathrm{rank}(\mathbf{V}_n^{\star})=1$, and the vector spanning $\mathrm{null}(\mathbf{Y}_n)$ is the principal generalized eigenvector because $\mathbf{Y}_n\mathbf{u}=\mathbf{0}$ reads $\mathbf{S}_n\mathbf{u}=\mathbf{B}_n\mathbf{u}$. $\blacksquare$

The eigenvalue one is simple in two common cases, namely when the $n$-th ST has a single active path and when the worst-case channels of all active paths are collinear, because in both cases $\mathbf{S}_n$ has rank one and Lemma~\ref{lem:nullity} already gives $\dim\mathrm{null}(\mathbf{Y}_n)\le1$. Driven by these structural insights, we adopt Proposition~\ref{prop:rank1} as the tightness certificate for our relaxed problem, which can be verified directly from the numerical KKT solution. When this certificate holds, the solver guarantees an exact rank-one outcome. In cases where the condition of Proposition~\ref{prop:rank1} fails, we explicitly recover a rank-one solution via a two-stage protocol. A targeted deterministic reduction—scaling the principal eigenvector of $\mathbf{V}_n^{\star}$ to the active QoS boundaries—serves as the primary mechanism. Gaussian randomization is relegated to a supplementary final stage, ensuring that this explicit hierarchy replaces the earlier heuristic reliance on random sampling.
\subsection{Standard Form and Variable Layout}
To implement the ADMM algorithm, we first map our matrix optimization variables into a standard real-valued vector form. Since the set of $M_t \times M_t$ Hermitian matrices, $\mathbb{H}^{M_t}$, naturally forms a real vector space of dimension $M_t^2$, we introduce the customized vectorization mapping $\mathrm{svec}: \mathbb{H}^{M_t} \to \mathbb{R}^{M_t^2}$, defined as:
\begin{equation}\label{eq:svec}
\begin{split}
  \mathrm{svec}(\mathbf{V})=&\Bigl[\,V_{11},\dots,V_{M_tM_t},\;
  \sqrt{2}\Re\{V_{12}\},\sqrt{2}\Im\{V_{12}\},\dots,\\
  &\sqrt{2}\Re\{V_{M_t-1,M_t}\},\sqrt{2}\Im\{V_{M_t-1,M_t}\}\,\Bigr]^{\mathsf{T}},
\end{split}
\end{equation}
where the first $M_t$ entries are the real diagonal elements and the remaining $M_t(M_t-1)$ entries are the real and imaginary parts of the strictly upper-triangular elements, each scaled by $\sqrt{2}$. With this scaling, $\mathrm{svec}$ preserves inner products and norms:
\begin{equation}\label{eq:isometry}
\begin{split}
  &\bigl\|\mathrm{svec}(\mathbf{V})\bigr\|_2^{2}
  =\sum_{k}V_{kk}^{2}+2\sum_{k<l}|V_{kl}|^{2}=\|\mathbf{V}\|_F^{2},\\
  &\bigl\langle\mathrm{svec}(\mathbf{V}),\mathrm{svec}(\mathbf{W})\bigr\rangle=\Re\bigl\{\mathrm{Tr}(\mathbf{V}\mathbf{W})\bigr\}.
\end{split}
\end{equation}

This property lets the $\mathbf{x}$-update of Section IV-D be computed by an eigenvalue decomposition in the matrix domain while the rest of the algorithm works with real vectors. We write $\mathrm{smat}=\mathrm{svec}^{-1}$.

All variables are stacked into a single vector $\mathbf{x}\in\mathbb{R}^{n_v}$: first the $N$ beamforming blocks, then one slack block per path. Paths are numbered in a flat order $t=1,\dots,P$ with $P=\sum_{n}P_n$; $o(t)$ is the user that owns path $t$, and each $(n,p)$ subscript is identified with its flat counterpart $t$. Table~\ref{tab:sec4_layout} gives the layout.
\begin{table}[htbp]
\centering
\caption{VARIABLE LAYOUT OF $\mathbf{x}\in\mathbb{R}^{n_v}$}
\label{tab:sec4_layout}
\begin{tabular}{lcccc}
\toprule
Block & Symbol & Content & Length & Cone \\
\midrule
$1\leq n\leq N$ & $\mathbf{x}_{V_n}$ & $\mathrm{svec}(\mathbf{V}_n)$ & $M_t^{2}$ & $\mathbb{S}_+^{M_t}$ \\
path $t$, part 1 & $\chi_t$ & SVC dual & $1$ & $\mathbb{R}_+$ \\
path $t$, part 2 & $\bm{\omega}_t$ & $[\bm{\omega}_{1,t};\dots;\bm{\omega}_{K_t,t}]$ & $K_tD$ & $\mathbb{R}_+^{K_tD}$ \\
path $t$, part 3 & $\bm{\psi}_t$ & $[\bm{\psi}_{1,t};\dots;\bm{\psi}_{K_t,t}]$ & $K_tD$ & $\mathbb{R}_+^{K_tD}$ \\
path $t$, part 4 & $s_t$ & slack of \eqref{eq:rc-a} & $1$ & $\mathbb{R}_+$ \\
\bottomrule
\end{tabular}
\end{table}
\noindent where $s_t$ is the slack variable that turns \eqref{eq:rc-a} into an equality; $K_t=|\mathcal{F}_t|$ is the number of support vectors of path $t$; and $M_t$ is the number of satellite transmit antennas defined in Section \ref{sc:3}, which is independent of the path index $t$.
Here $\mathbb{S}_+^{M_t}$ denotes the cone of $M_t\times M_t$ positive semidefinite matrices and $\mathbb{R}_+^{m}$ the nonnegative orthant of dimension $m$. The offsets are $\mathrm{off}_{V_n}=(n-1)M_t^{2}$ and $\mathrm{off}_{t}=NM_t^{2}+\sum_{t'<t}(2+2K_{t'}D)$, and the total dimension is
\begin{equation}\label{eq:nv}
  n_v=NM_t^{2}+\sum_{t=1}^{P}\bigl(2+2K_tD\bigr).
\end{equation}

The cone $\mathcal{K}$ is the product of the $N$ semidefinite cones $\mathbb{S}_+^{M_t}$ and the nonnegative orthant $\mathbb{R}_+^{n_v-NM_t^{2}}$ for the remaining entries.
\begin{equation}\label{eq:coneK}
\begin{split}
\mathcal{K}=\Bigl\{ \mathbf{x}\,\big|\,
& \mathrm{smat}(\mathbf{x}_{V_n})\succeq\mathbf{0},\quad n=1,\dots,N; \chi_t\geq 0,\\
&  \bm{\omega}_t\geq \mathbf{0}, \bm{\psi}_t\geq \mathbf{0}, s_t\geq 0, t=1,\dots,P \Bigr\}.
\end{split}
\end{equation}
where the first condition means that each beamforming block $\mathbf{x}_{V_n}$ reconstructs a positive semidefinite matrix, and the remaining inequalities are entrywise.

Introducing the slack $s_t\geq 0$ turns the inequality \eqref{eq:rc-a} into an equality. For each path $t$, owned by user $o(t)$ and with interference set $\mathcal{I}_t$, the three constraint groups are
\begin{subequations}\label{eq:con}
\begin{align}
  &2\,\mathcal{T}(\mathbf{A}_t)\check{\bar{\mathbf{g}}}_t
  +\mathbf{Q}_t\sum_{i=1}^{K_t}\bigl(\bm{\omega}_{i,t}-\bm{\psi}_{i,t}\bigr)=\mathbf{0},\label{eq:con-a}\\
  &\check{\bar{\mathbf{g}}}_t^{\mathsf{T}}\mathcal{T}(\mathbf{A}_t)\check{\bar{\mathbf{g}}}_t
  \!+\!\Gamma_t\rho_t^{2}\sum_{m\in\mathcal{I}_t}\mathrm{Tr}(\mathbf{V}_m)
  \!+\!\sum_{i=1}^{K_t}\bigl(\bm{\omega}_{i,t}-\bm{\psi}_{i,t}\bigr)^{\mathsf{T}}\mathbf{Q}_t\bm{\xi}_{i,t}\nonumber\\
  &+\varrho_t\chi_t+s_t=-\Gamma_t\sigma^{2},\label{eq:con-b}\\
  &\bm{\omega}_{i,t}+\bm{\psi}_{i,t}-\lambda_{i,t}\chi_t\mathbf{1}_D=\mathbf{0},\quad i=1,\dots,K_t,\label{eq:con-c}
\end{align}
\end{subequations}
where $\mathbf{A}_t=\mathbf{V}_{o(t)}-\Gamma_t\sum_{m\in\mathcal{I}_t}\mathbf{V}_m$ and the three groups come from \eqref{eq:rc-b}, \eqref{eq:rc-a}, and \eqref{eq:rc-c}, respectively. Stacking \eqref{eq:con-a}--\eqref{eq:con-c} path by path gives
\begin{equation}\label{eq:consys}
  \mathcal{A}\mathbf{x}=\mathbf{b},\quad \mathcal{A}\in\mathbb{R}^{n_c\times n_v},\quad
  n_c=\sum_{t=1}^{P}\bigl(1+(1+K_t)D\bigr),
\end{equation}
where $\mathbf{b}$ has the entry $-\Gamma_t\sigma^{2}$ in the row of \eqref{eq:con-b} for each path and zeros elsewhere. Table~\ref{tab:sec4_rows} summarizes the row layout.
\begin{table}[H]
\centering
\caption{CONSTRAINT ROW LAYOUT (PER PATH $t$)}
\label{tab:sec4_rows}
\setlength{\tabcolsep}{3pt} 
\renewcommand{\tabularxcolumn}[1]{m{#1}}
\begin{tabularx}{\columnwidth}{X c c c} 
\toprule
Rows & Source & Count & RHS \\
\midrule
$1,\dots,D$ & \eqref{eq:con-a}, gradient matching & $D=2M_t$ & $0$ \\
$(D+1)$ & \eqref{eq:con-b}, worst-case value + slack & $1$ & $-\Gamma_t\sigma^{2}$ \\
$D+2,\dots,\newline (D+1)+K_tD$ & \eqref{eq:con-c}, SVC dual coupling & $K_tD$ & $0$ \\
\bottomrule
\end{tabularx}
\end{table}

Since $\mathrm{Tr}(\mathbf{V}_n)$ is the sum of the first $M_t$ entries of $\mathrm{svec}(\mathbf{V}_n)$, the objective is $\mathbf{c}^{\mathsf{T}}\mathbf{x}$ with $\mathbf{c}\in\{0,1\}^{n_v}$ having ones exactly at the diagonal positions of the $N$ beamforming blocks. Problem \eqref{eq:P1} becomes the conic program
\begin{subequations}\label{eq:P2}
  \begin{align}
    &\min_{\mathbf{x}}\;\mathbf{c}^{\mathsf{T}}\mathbf{x} \\
    \text{s.t.}\quad &\mathcal{A}\mathbf{x}=\mathbf{b},\mathbf{x}\in\mathcal{K}.
  \end{align}
\end{subequations}
where the matrix $\mathcal{A}$ is built entirely offline to eliminate redundant online computations. Each column $j$ of $\mathcal{A}$ directly corresponds to the image of the basis vector $\mathbf{e}_j$ under the linear constraints \eqref{eq:con-a}--\eqref{eq:con-c}.
\subsection{GPU-Accelerated ADMM Solution}
\begin{figure*}[htbp]
    \centering
    \includegraphics[width=2\columnwidth]{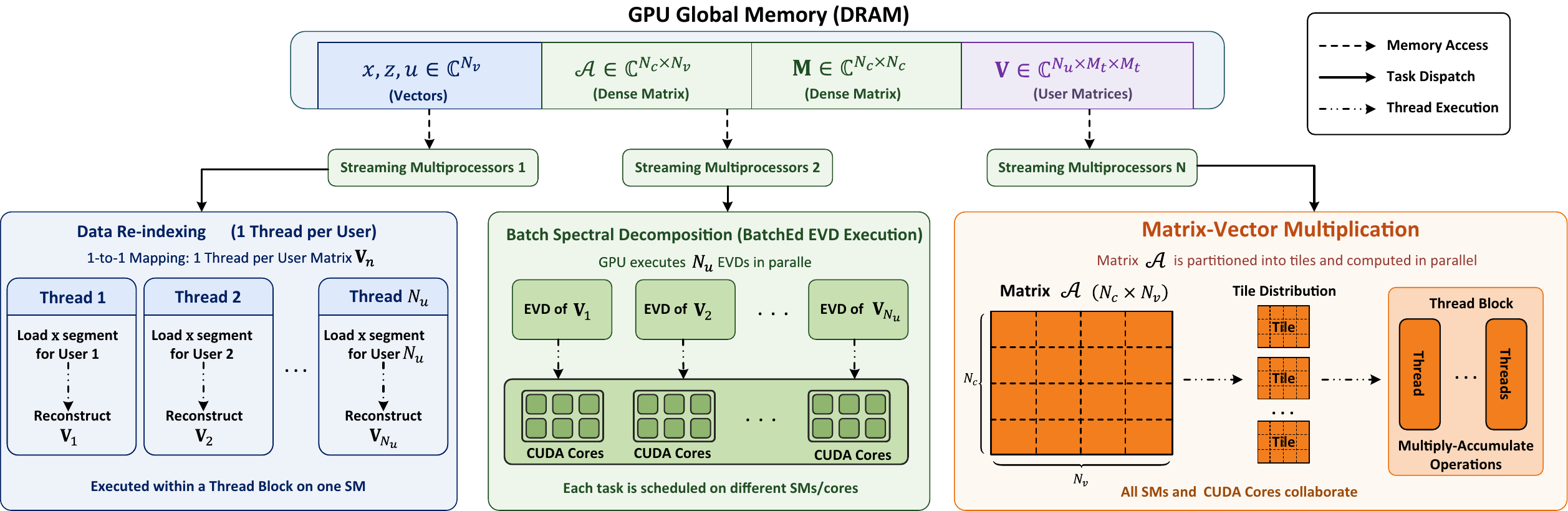}
    \caption{Thread-level scheduling and task assignment workflow for the GPU-accelerated ADMM framework.}
    \label{fig:gpu_architecture}
\end{figure*}
The primary computational bottleneck in solving \eqref{eq:P2} lies in the simultaneous enforcement of the positive semidefinite cone and the linear equations. We achieve this by introducing a duplicated consensus variable, thereby splitting the complex problem into two tractable parts—a cone portion and an affine portion:
\begin{subequations}\label{eq:split}
\begin{align}
&\min_{\mathbf{x},\mathbf{z}}  f(\mathbf{x})+g(\mathbf{z}) \\
\text{s.t.}\quad & \mathbf{x}-\mathbf{z}=\mathbf{0}.
\end{align}
\end{subequations}
\begin{equation}\label{eq:splitfuncs}
  f(\mathbf{x})=\mathbf{c}^{\mathsf{T}}\mathbf{x}+\mathbb{I}_{\mathcal{K}}(\mathbf{x}),\,
  g(\mathbf{z})=\mathbb{I}_{\mathcal{L}}(\mathbf{z}),\,
  \mathcal{L}\triangleq\{\mathbf{z}|\mathcal{A}\mathbf{z}=\mathbf{b}\},
\end{equation}
where $\mathbb{I}_{\mathcal{X}}(\cdot)$ is the indicator function of the set $\mathcal{X}$ (zero on $\mathcal{X}$, $+\infty$ elsewhere).

The augmented Lagrangian in scaled form, with penalty $\beta>0$ and scaled dual variable $\mathbf{u}=\mathbf{y}_{L_\mathrm{ADMM}}/\beta$, is
\begin{equation}\label{eq:auglag}
  L_{\mathrm{ADMM}}(\mathbf{x},\mathbf{z},\mathbf{u})=f(\mathbf{x})+g(\mathbf{z})
  +\frac{\beta}{2}\bigl\|\mathbf{x}-\mathbf{z}+\mathbf{u}\bigr\|_2^{2}-\frac{\beta}{2}\|\mathbf{u}\|_2^{2},
\end{equation}
where $\mathbf{y}_{L_\mathrm{ADMM}}$ is the unscaled dual variable. With over-relaxation parameter $\alpha\in(0,2)$, one ADMM iteration is
\begin{subequations}\label{eq:admm}
\begin{align}
  \mathbf{x}^{k+1}&=\arg\min_{\mathbf{x}}\;f(\mathbf{x})+\frac{\beta}{2}\bigl\|\mathbf{x}-\mathbf{z}^{k}+\mathbf{u}^{k}\bigr\|_2^{2},\label{eq:admm-a}\\
  \hat{\mathbf{x}}^{k+1}&=\alpha\mathbf{x}^{k+1}+(1-\alpha)\mathbf{z}^{k},\label{eq:admm-b}\\
  \mathbf{z}^{k+1}&=\arg\min_{\mathbf{z}}\;g(\mathbf{z})+\frac{\beta}{2}\bigl\|\hat{\mathbf{x}}^{k+1}-\mathbf{z}+\mathbf{u}^{k}\bigr\|_2^{2},\label{eq:admm-c}\\
  \mathbf{u}^{k+1}&=\mathbf{u}^{k}+\hat{\mathbf{x}}^{k+1}-\mathbf{z}^{k+1},\label{eq:admm-d}
\end{align}
\end{subequations}

Since $\mathcal{K}$ is a product cone and $\mathbf{c}$ has support only on the beamforming blocks, the $\mathbf{x}$-update \eqref{eq:admm-a} decouples into $N$ matrix problems and one element-wise problem. Let $\mathbf{w}=\mathbf{z}^{k}-\mathbf{u}^{k}$, which is partitioned into the blocks $\mathbf{w}_{V_n}$ for $n=1,\dots,N$, and $\mathbf{w}_{\mathrm{slack}}$ corresponding to parts 1-4 according to Table~\ref{tab:sec4_layout}. For the beamforming block, the subproblem can be derived by rewriting $\mathrm{Tr}(\mathbf{V})=\langle\mathbf{I},\mathbf{V}\rangle$ and completing the square:
\begin{equation}\label{eq:projbeam}
  \arg\min_{\mathbf{V}\succeq\mathbf{0}}\;\mathrm{Tr}(\mathbf{V})+\frac{\beta}{2}\|\mathbf{V}-\mathbf{W}\|_F^{2}=\Pi_{\mathbb{S}_+}\Bigl(\mathbf{W}-\frac{1}{\beta}\mathbf{I}\Bigr),
\end{equation}
where $\mathbf{W}=\mathrm{smat}(\mathbf{w}_{V_n})\in\mathbb{H}^{M_t}$ is the matrix form of the $n$-th beamforming block of $\mathbf{w}$. For any eigenvalue decomposition $\mathbf{M}=\mathbf{U}\bm{\Lambda}\mathbf{U}^{H}$, the projection onto $\mathbb{S}_+$ is defined as $\Pi_{\mathbb{S}_+}(\mathbf{M})=\mathbf{U}\max(\bm{\Lambda},\mathbf{0})\mathbf{U}^{H}$. The identity holds for any $\mathbf{W}\in\mathbb{H}^{M_t}$ and $\beta>0$ and represents the projection of $\mathbf{W}-\beta^{-1}\mathbf{I}$ onto $\mathbb{S}_+$.

On the slack entries $(\chi_t,\bm{\omega}_t,\bm{\psi}_t,s_t)$ since $\mathbf{c}$ is supported only on the beamforming blocks, and the cone reduces to the nonnegative orthant, so that the quadratic penalty decouples entry by entry and each entry is simply clipped at zero. Therefore the whole $\mathbf{x}$-update is
\begin{equation}\label{eq:xupdate}
\begin{split}
  &\mathbf{x}^{k+1}_{V_n}=\mathrm{svec}\Bigl(\Pi_{\mathbb{S}_+}\bigl(\mathrm{smat}(\mathbf{w}_{V_n})-\beta^{-1}\mathbf{I}\bigr)\Bigr),\\
  &\mathbf{x}^{k+1}_{\mathrm{slack}}=\max(\mathbf{w}_{\mathrm{slack}},\mathbf{0}),
\end{split}
\end{equation}

Let $\mathbf{t}=\hat{\mathbf{x}}^{k+1}+\mathbf{u}^{k}$. The $\mathbf{z}$-update Eq. \eqref{eq:admm-c} is equivalent to solves $\min_{\mathbf{z}}\frac12\|\mathbf{z}-\mathbf{t}\|_2^{2}$ subject to $\mathcal{A}\mathbf{z}=\mathbf{b}$. Its Lagrangian is $L_{\mathrm{up}}=\frac12\|\mathbf{z}-\mathbf{t}\|^{2}+\mathbf{y}_{L_{\mathrm{up}}}^{\mathsf{T}}(\mathcal{A}\mathbf{z}-\mathbf{b})$; the stationarity condition gives $\mathbf{z}=\mathbf{t}-\mathcal{A}^{\mathsf{T}}\mathbf{y}_{L_{\mathrm{up}}}$, and substituting into $\mathcal{A}\mathbf{z}=\mathbf{b}$ yields $\mathcal{A}\mathbf{t}-\mathcal{A}\mathcal{A}^{\mathsf{T}}\mathbf{y}_{L_{\mathrm{up}}}=\mathbf{b}$. Therefore the $\mathbf{z}$-update is
\begin{equation}\label{eq:zproj}
  \mathbf{z}^{k+1}=\mathbf{t}-\mathcal{A}^{\mathsf{T}}
  \mathbf{M}\bigl(\mathcal{A}\mathbf{t}-\mathbf{b}\bigr),
\end{equation}
where $\mathbf{M}=(\mathcal{A}\mathcal{A}^{\mathsf{T}})^{-1}$ is only computed once offline.

The primal and dual residuals are
\begin{equation}\label{eq:residuals}
  r^{k}=\bigl\|\mathbf{x}^{k}-\mathbf{z}^{k}\bigr\|_2,\qquad
  s^{k}=\beta\bigl\|\mathbf{z}^{k}-\mathbf{z}^{k-1}\bigr\|_2,
\end{equation}
where $r^{k}$ measures the mismatch between the two copies of the variable and $s^{k}$ how much the solution changes between iterations. The algorithm stops when
\begin{equation}\label{eq:stopping}
\begin{split}
  &r^{k}\leq\sqrt{n_v}\,\epsilon_{\mathrm{abs}}+\epsilon_{\mathrm{rel}}\max\bigl\{\|\mathbf{x}^{k}\|_2,\|\mathbf{z}^{k}\|_2\bigr\},\\
  &s^{k}\leq\sqrt{n_v}\,\epsilon_{\mathrm{abs}}+\epsilon_{\mathrm{rel}}\,\beta\|\mathbf{u}^{k}\|_2,
\end{split}
\end{equation}
where $\epsilon_{\mathrm{abs}}$ and $\epsilon_{\mathrm{rel}}$ are the absolute and relative tolerances. To balance the two residuals, the penalty is adapted every $k_a$ iterations by
\begin{equation}\label{eq:adapt}
  \beta\leftarrow
  \begin{cases}
    \tau\beta,\;\mathbf{u}\leftarrow\mathbf{u}/\tau, & r^{k}>\eta\,s^{k},\\
    \beta/\tau,\;\mathbf{u}\leftarrow\tau\mathbf{u}, & s^{k}>\eta\,r^{k},\\
    \beta, & \text{otherwise},
  \end{cases}
\end{equation}
with $\tau=1.5$ and $\eta=10$; the rescaling of $\mathbf{u}$ is needed because $\mathbf{u}=\mathbf{y}/\beta$, so changing $\beta$ must not change the unscaled dual variable $\mathbf{y}$. The penalty is held constant after iteration $k_f$ to keep the iteration stable. We initialize $\beta_0=1/(\Gamma_{n,p}\sigma^{2})$, which matches the scale of the right-hand side of \eqref{eq:con-b}.

Algorithm~\ref{alg:assembly} summarizes the offline stage and Algorithm~\ref{alg:gpuadmm} the device-resident iteration.
\begin{algorithm}[h]
\caption{Offline Set Learning and Operator Assembly (Host/CPU)}
\label{alg:assembly}
\begin{algorithmic}[1]
\REQUIRE CSI samples $\{\mathbf{g}^{(s)}_{n,p}\}$, nominal channels $\bar{\mathbf{g}}_{n,p}$, thresholds $\Gamma_{n,p}$, noise power $\sigma^{2}$, outage $\epsilon$, interference sets $\mathcal{I}(n,p)$
\STATE \textbf{for} each path $t=(n,p)$ \textbf{do}
\STATE \quad embed the samples by \eqref{eq:samples}; compute $\bm{\Sigma}_t$ and $\mathbf{Q}_t=\bm{\Sigma}_t^{-1/2}$
\STATE \quad solve the SVC dual \eqref{eq:svcdual} with the kernel \eqref{eq:kernel}; obtain $\lambda_{s,t}$, $\mathcal{F}_t$, $\mathcal{B}_t$, and $K_t=|\mathcal{F}_t|$
\STATE \quad compute $\varrho_t$ by \eqref{eq:l1radius} and $\hat{\rho}_t$ by \eqref{eq:rhobound}
\STATE \textbf{end for}
\STATE build the layout of Table~\ref{tab:sec4_layout}; set $n_v$ by \eqref{eq:nv} and $n_c$ by \eqref{eq:consys}
\STATE \textbf{for} $j=1$ \textbf{to} $n_v$ \textbf{do} set column $j$ of $\mathcal{A}$ to the image of $\mathbf{e}_j$ under \eqref{eq:con-a}--\eqref{eq:con-c} \textbf{end for}
\STATE assemble $\mathbf{b}$ and $\mathbf{c}$; form $\mathbf{M}=(\mathcal{A}\mathcal{A}^{\mathsf{T}})^{-1}$
\STATE transfer $\mathcal{A}$, $\mathbf{M}$, $\mathbf{b}$, $\mathbf{c}$ to device memory
\ENSURE $\mathcal{A}$, $\mathbf{M}$, $\mathbf{b}$, $\mathbf{c}$ on the GPU
\end{algorithmic}
\end{algorithm}
\begin{algorithm}[h]
\caption{GPU-Accelerated ADMM for Problem \eqref{eq:P2}}
\label{alg:gpuadmm}
\begin{algorithmic}[1]
\REQUIRE $\mathcal{A}$, $\mathbf{M}$, $\mathbf{b}$, $\mathbf{c}$; parameters $\beta_0$, $\alpha$, $\tau$, $\eta$, $k_a$, $k_f$, $\epsilon_{\mathrm{abs}}$, $\epsilon_{\mathrm{rel}}$, $k_{\max}$
\STATE initialize $\mathbf{x}\leftarrow\mathbf{0}$, $\mathbf{z}\leftarrow\mathbf{0}$, $\mathbf{z}^{\mathrm{old}}\leftarrow\mathbf{0}$, $\mathbf{u}\leftarrow\mathbf{0}$, $\beta\leftarrow\beta_0$
\FOR{$k=1$ \textbf{to} $k_{\max}$}
\STATE $\mathbf{w}\leftarrow\mathbf{z}-\mathbf{u}$
\FOR{$n=1$ \textbf{to} $N$ \textbf{in parallel}}
\STATE \quad $\mathbf{W}\leftarrow\mathrm{smat}(\mathbf{w}_{V_n})$; compute the eigendecomposition $\mathbf{W}-\beta^{-1}\mathbf{I}=\mathbf{U}\bm{\Lambda}\mathbf{U}^{H}$
\STATE \quad set $\mathbf{x}_{V_n}\leftarrow\mathrm{svec}\bigl(\mathbf{U}\max(\bm{\Lambda},\mathbf{0})\mathbf{U}^{H}\bigr)$ \hfill \emph{\eqref{eq:xupdate}}
\ENDFOR
\STATE $\mathbf{x}_{\mathrm{slack}}\leftarrow\max(\mathbf{w}_{\mathrm{slack}},\mathbf{0})$ \hfill \emph{\eqref{eq:xupdate}}
\STATE $\hat{\mathbf{x}}\leftarrow\alpha\mathbf{x}+(1-\alpha)\mathbf{z}$ \hfill \emph{\eqref{eq:admm-b}}
\STATE $\mathbf{v}\leftarrow\hat{\mathbf{x}}+\mathbf{u}$
\STATE $\mathbf{z}\leftarrow\mathbf{v}-\mathcal{A}^{\mathsf{T}}\,\mathbf{M}(\mathcal{A}\mathbf{v}-\mathbf{b})$ \hfill \emph{\eqref{eq:zproj}}
\STATE $\mathbf{u}\leftarrow\mathbf{u}+\hat{\mathbf{x}}-\mathbf{z}$ \hfill \emph{\eqref{eq:admm-d}}
\STATE $r\leftarrow\|\mathbf{x}-\mathbf{z}\|_2$; $s\leftarrow\beta\|\mathbf{z}-\mathbf{z}^{\mathrm{old}}\|_2$; $\mathbf{z}^{\mathrm{old}}\leftarrow\mathbf{z}$ \hfill \emph{\eqref{eq:residuals}}
\IF{\eqref{eq:stopping} holds}
\STATE \textbf{break}
\ENDIF
\IF{$k<k_f$ \textbf{and} $k\bmod k_a=0$}
\STATE update $\beta$ and rescale $\mathbf{u}$ by \eqref{eq:adapt}
\ENDIF
\ENDFOR
\STATE recover $\mathbf{V}_n\leftarrow\mathrm{smat}(\mathbf{x}_{V_n})$ for $n=1,\dots,N$
\ENSURE beamforming matrices $\{\mathbf{V}_n\}$
\end{algorithmic}
\end{algorithm}

The beamforming update \eqref{eq:xupdate} needs $N$ Hermitian eigendecompositions of size $M_t\times M_t$, costing $O(NM_t^{3})$ operations, plus $O(NM_t^{3})$ to rebuild $\mathbf{U}\max(\bm{\Lambda},\mathbf{0})\mathbf{U}^{H}$; the projection \eqref{eq:zproj} needs two products with $\mathcal{A}$ and one with $\mathbf{M}$, i.e. $O(2n_cn_v+n_c^{2})$ operations; the remaining steps are element-wise and cost $O(n_v)$. The total per-iteration cost is $O\bigl(n_cn_v+n_c^{2}+NM_t^{3}\bigr)$, dominated by the two dense matrix--vector products, which are memory-bandwidth bound and therefore well suited to a GPU. Before the iterations, the matrix $\mathcal{A}$ is assembled once at cost $O(n_vn_c)$, and $\mathcal{A}\mathcal{A}^{\mathsf{T}}$ is then formed and factorized once at cost $O(n_c^{2}n_v+n_c^{3})$. With $\bar{K}$ the average number of support vectors per path, \eqref{eq:nv} and \eqref{eq:consys} become $n_v\approx NM_t^{2}+4P\bar{K}M_t$ and $n_c\approx 2P(1+\bar{K})M_t$, so $n_c<n_v$ in typical systems and the matrix inverted in \eqref{eq:zproj} is only of size $n_c\times n_c$.


With the tractable robust optimization framework and its efficient ADMM-based solver fully developed, we now proceed to evaluate its real-world performance. Section V will provide extensive simulation results to validate the robustness, energy efficiency, and computational superiority of the proposed method.

\section{\uppercase{{\large S}imulation {\large R}esults}}
\begin{table}[H]
\centering
\caption{SIMULATION PARAMETERS}
\label{tab:sim_params}
\begin{tabular}{lll}
\toprule
Parameter & Symbol & Value \\
\midrule
Carrier frequency & $f_c$ & 2 GHz \\
Orbital altitude & $h$ & 500 km \\
Satellite orbital velocity & $v_s$ & 7.62 km/s \\
Subcarrier spacing / bandwidth & $\Delta f$, $B$ & 156 kHz, 20 MHz \\
Delay and Doppler bins & $N_\tau$, $N_\nu$ & 128, 256 \\
Sampling period / CP length & $T_s$, $L_{\mathrm{cp}}$ & 0.05 $\mu$s, 32 \\
Antenna spacing (ULA) & $d$ & $\lambda/2$ \\
Paths per ST & $P_n$ & 2 \\
Angular resolution & $\Delta\Theta_{\min}$ & $2\lambda/(M_t d)$ \\
Outage probability & $\epsilon$ & 0.1 \\
Noise power & $\sigma^2$ & 0.1 \\
SVC penalty constant & $C$ & $P_n/(\epsilon S)$ \\
ADMM initial penalty & $\beta_0$ & $1/(\Gamma\sigma^2)$ \\
ADMM relaxation / penalty & $\alpha$, $\eta$, $\tau$ & 1.6, 10, 1.5 \\
ADMM adaptation period & $k_a$ & 50 iterations \\
ADMM max / min iterations & $k_{\max}$, $k_{\min}$ & 50000 / 200 \\
ADMM stopping tolerances & $\varepsilon_{\mathrm{abs}}$, $\varepsilon_{\mathrm{rel}}$ & $10^{-8}$, $5\times10^{-4}$ \\
CPU platform & --- & AMD 7950X3D \\
GPU platform & --- & NVIDIA RTX 3080 \\
\bottomrule
\end{tabular}
\end{table}
In this section, a series of simulation experiments are conducted to evaluate the performance of the proposed robust optimization method. The key system parameters are listed in Table \ref{tab:sim_params}. The evaluation proceeds from the modeling layer to the computational layer. The uncertainty model is assessed first, where the SVC-based set is compared with the conventional Box and Ellipsoid sets and with the Average (AVG) and worst-case channel condition (WCCC) baselines that represent channel overestimation and underestimation, respectively, so that the impact of CSI errors on the transmit power and on the achieved reliability can be quantified on the same channel samples. The computational layer is assessed next, where the GPU-accelerated ADMM is benchmarked against the same algorithm executed on a CPU and against the interior-point solvers of CVX with SDPT3 and SeDuMi, as the antenna array and the number of served STs are scaled up. The results show that the proposed uncertainty model reduces the surplus transmit power of robust beamforming, and the proposed solver improves the computational efficiency as the sizes of antenna array and user population increase.
\begin{figure}[h]
    \centering
    \includegraphics[width=0.98\columnwidth]{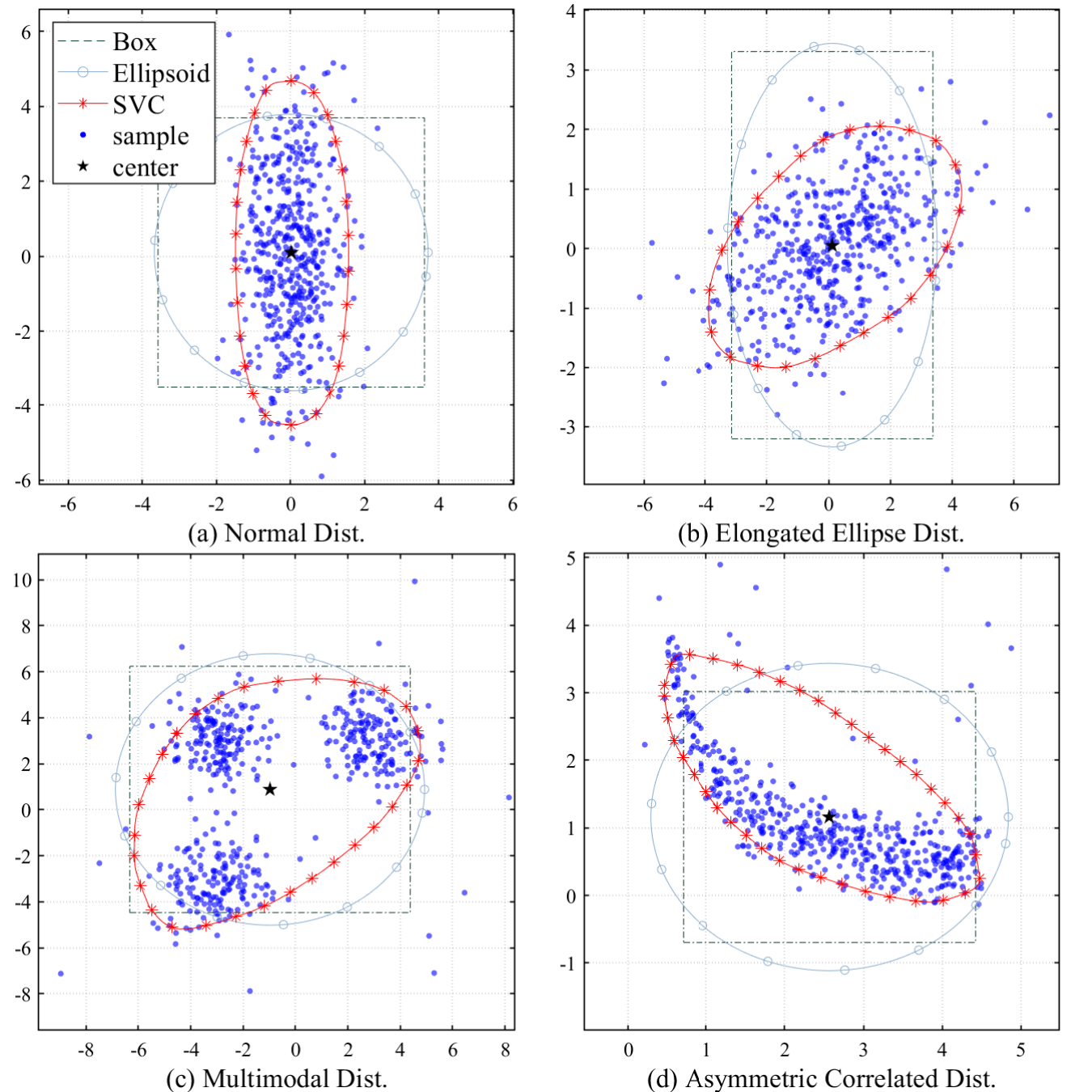}
    \caption{The uncertainty sets constructed on different datasets with $\epsilon = 0.1$ using the Box, Ellipsoid-based, and SVC approaches are marked in green, azure, and red, respectively.}  
    \label{fig:all_Dist} 
\end{figure}

\begin{figure}[htbp]
    \centering
    \includegraphics[width=0.8\columnwidth]{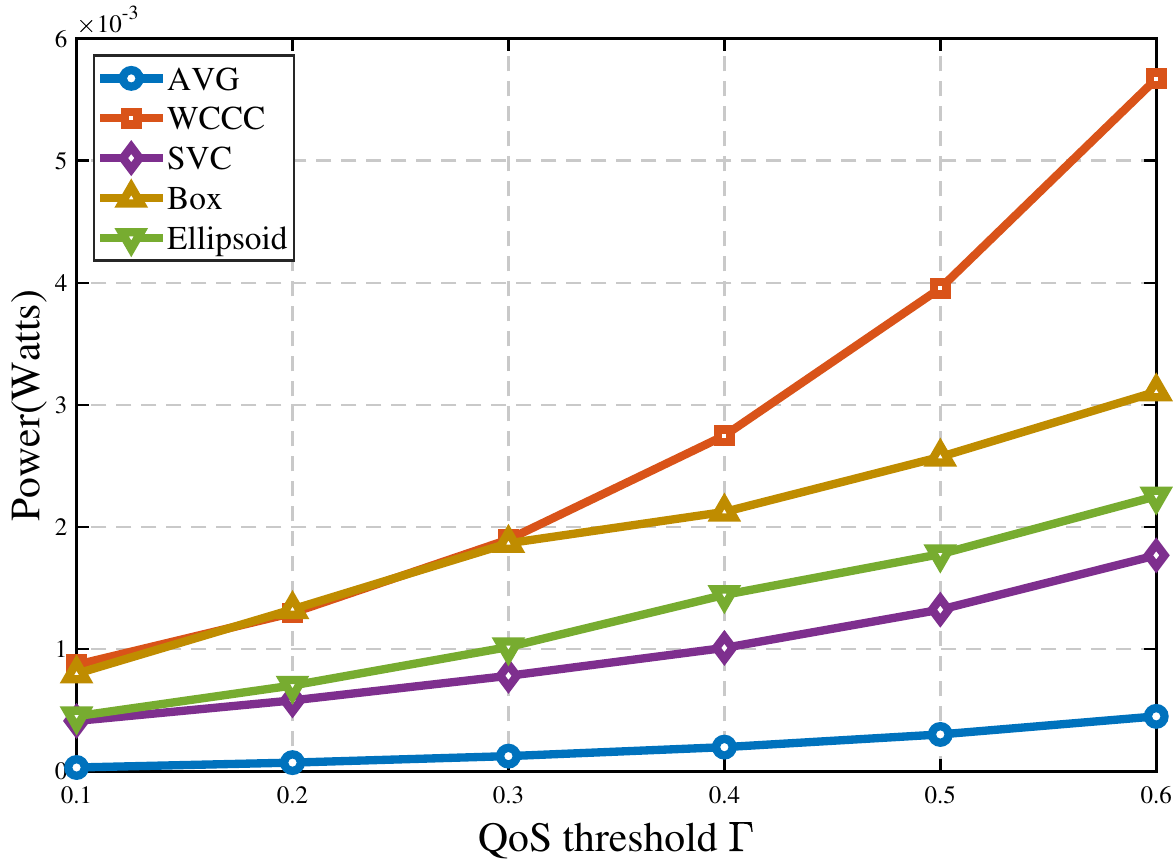}
    \caption{Total transmit power versus the predefined QoS thresholds $\Gamma$ under different methods.}
    \label{fig:power_vs_qos}
\end{figure}

\begin{figure}[htbp]
    \centering
    \includegraphics[width=0.8\columnwidth]{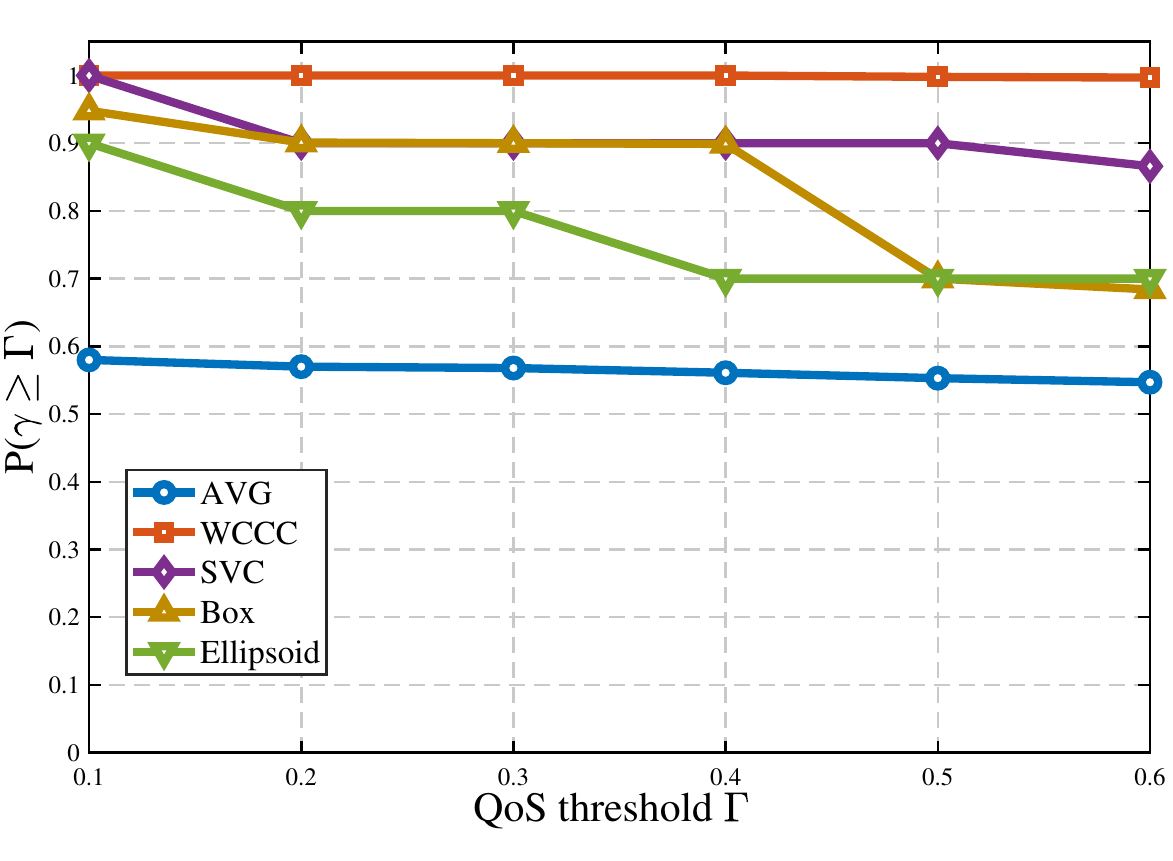}
    \caption{Achieved transmission reliability versus the predefined QoS thresholds $\Gamma$ under different methods.}
    \label{fig:reliability_vs_qos}
\end{figure}

Fig. \ref{fig:all_Dist} illustrates the uncertainty sets constructed on different
datasets with $\epsilon=0.1$ using the Box, Ellipsoid-based, and SVC approaches. The Box and Ellipsoid sets enclose a considerably larger volume than the SVC set and the results show that this extra volume lies in regions with no real samples. Being
symmetric, these two sets must extend along every direction in order to reach the
required coverage, including the directions that contain no samples at all, while the separated modes and the curved tail of the data are still only partially covered. The set produced by the SVC method is asymmetric, so it follows the shape of the data rather than a prescribed geometry and encloses the same fraction of samples within a much smaller volume. These geometric differences have implications for system performance. Specifically, the volume of an uncertainty set dictates the transmit power required by its associated beamforming design and the actual QoS for users hinges on whether the set successfully captures the true channel realizations. Figs. \ref{fig:power_vs_qos} and \ref{fig:reliability_vs_qos} quantify these impacts on power consumption and reliability, respectively.
\begin{figure}[htbp]
    \centering
    \includegraphics[width=0.8\columnwidth]{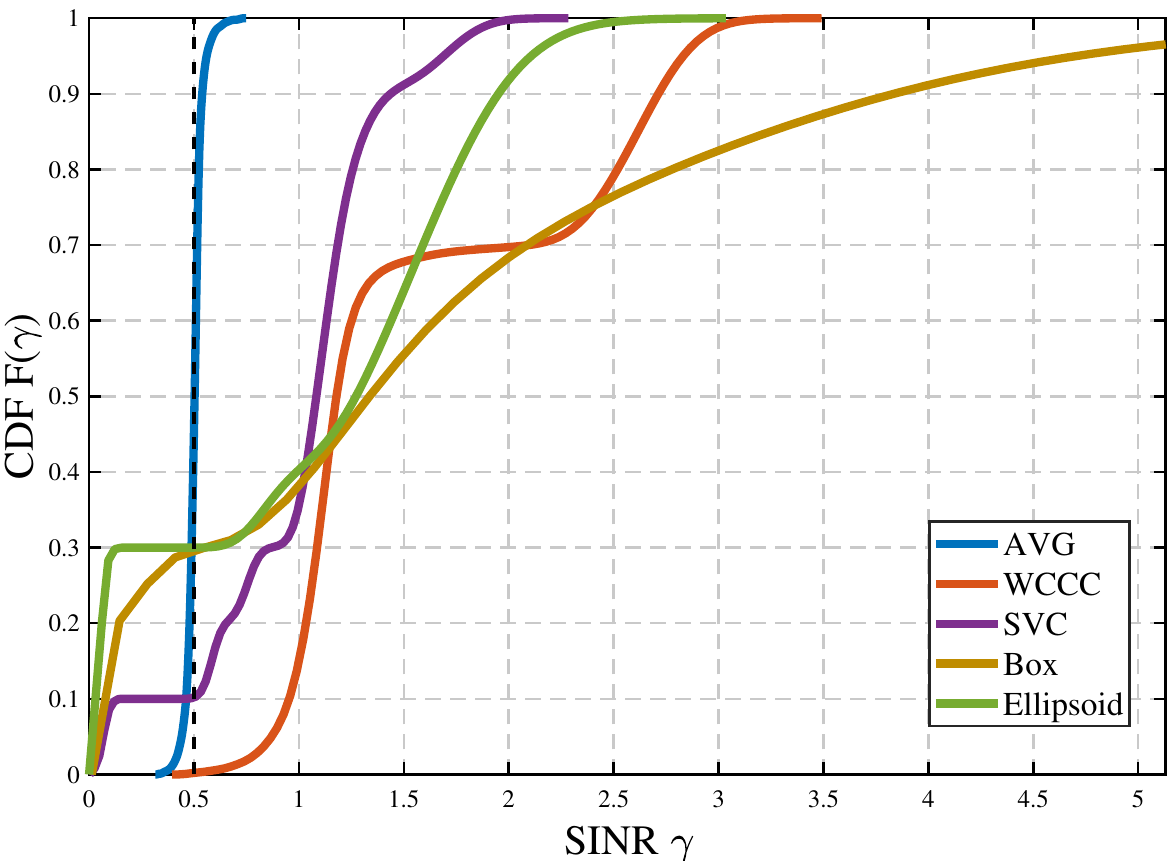}
    \caption{Cumulative distribution of the achievable SINR for various methods assuming QoS threshold $\Gamma = 0.5$.}
    \label{fig:cdf_sinr_qos05}
\end{figure}
\begin{figure}[htbp]
    \centering
    \includegraphics[width=0.8\columnwidth]{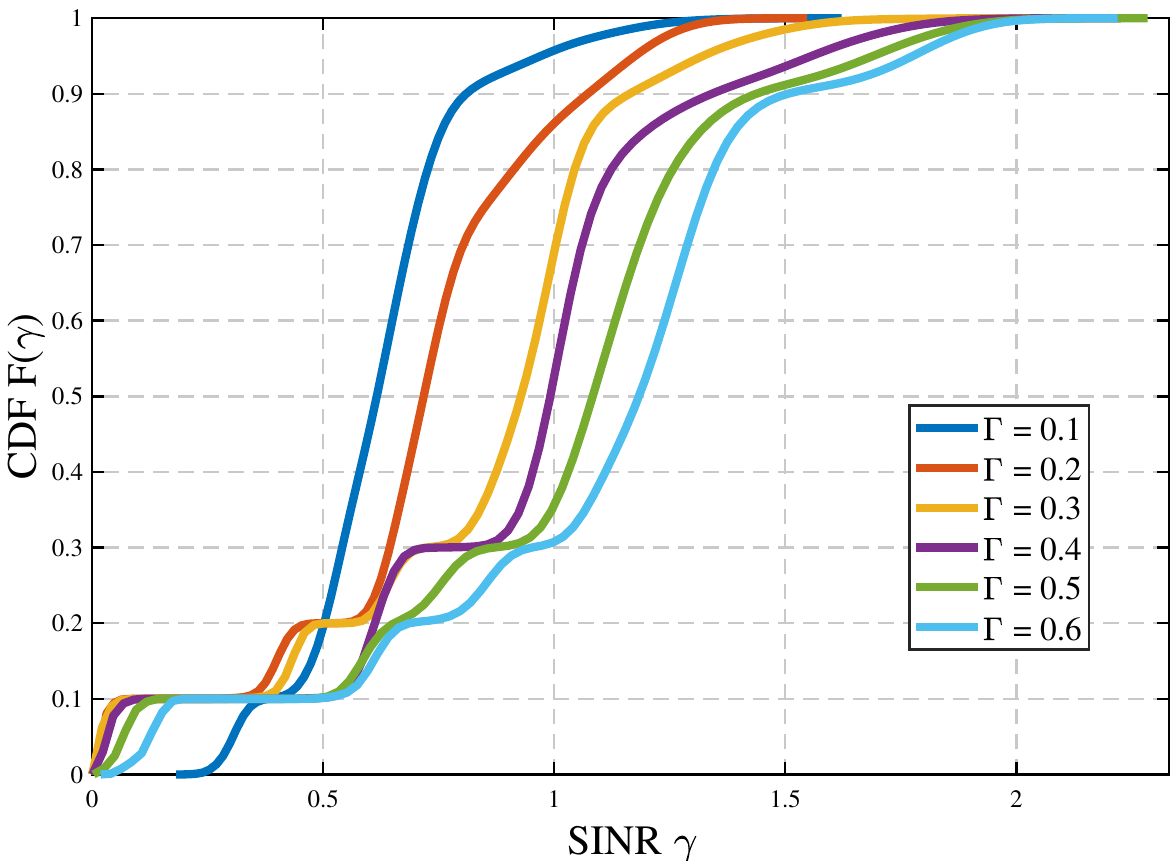}
    \caption{Cumulative distribution of the achievable SINR under the SVC-based method across varying predefined QoS thresholds $\Gamma$.}
    \label{fig:cdf_sinr_svc_vary_qos}
\end{figure}

Fig. \ref{fig:cdf_sinr_qos05} and Fig. \ref{fig:cdf_sinr_svc_vary_qos} present the cumulative distribution of achievable SINR for different schemes. The former corresponds to a fixed QoS threshold $\Gamma = 0.5$, while the latter illustrates the performance of the SVC-based method under various predefined QoS thresholds $\Gamma$. The AVG method simulates a scenario where the system is overly optimistic about the channel quality, leading to an aggressive beamforming strategy that fails to meet the SINR threshold for a large portion of the sample points. Conversely, the WCCC method simulates an overly pessimistic scenario, causing the SINR values for nearly all channel realizations to far exceed the threshold. The proposed SVC method provides a clear improvement by tracking the required threshold closely, reducing the resource waste associated with WCCC.

Fig. \ref{fig:power_vs_qos} and Fig. \ref{fig:reliability_vs_qos} depict the total transmit power and the achieved transmission reliability versus the predefined QoS thresholds $\Gamma$, respectively. As $\Gamma$ increases, the power required by all robust methods exhibits an upward trend. The WCCC-based method consumes the highest power because it is provisioned the worst-case CSI; the Box and Ellipsoid methods consume the next highest amounts; and the proposed SVC method consumes significantly less than the above three, saving up to $50\%$ of the transmit power compared with the Box model. This saving has a geometric origin, because to exclude the same fraction $\epsilon$ of the samples, the symmetric Box and Ellipsoid sets must be enlarged along every direction, including those that the true CSI never visits, so that a considerable part of the power they spend is devoted to directions that carry no channel energy. The reliability curves show that the AVG method consumes the least power but attains a reliability far below the prescribed level, which limits its applicability, whereas the SVC method attains $1-\epsilon$ over the whole range of $\Gamma$. Notably, the larger volume of the Box and Ellipsoid sets does not translate into a better guarantee: as observed in Fig. \ref{fig:all_Dist}, these sets still leave the asymmetric and multimodal regions of the CSI distribution only partially covered, so their achieved reliability remains below the prescribed level even though they consume more power. The SVC set is fitted to the shape of the data and therefore avoids both the surplus margin and the coverage gap, enabling it to achieve the prescribed reliability with the lowest transmit power.
\begin{figure}[htbp]
    \centering
    \includegraphics[width=0.9\columnwidth]{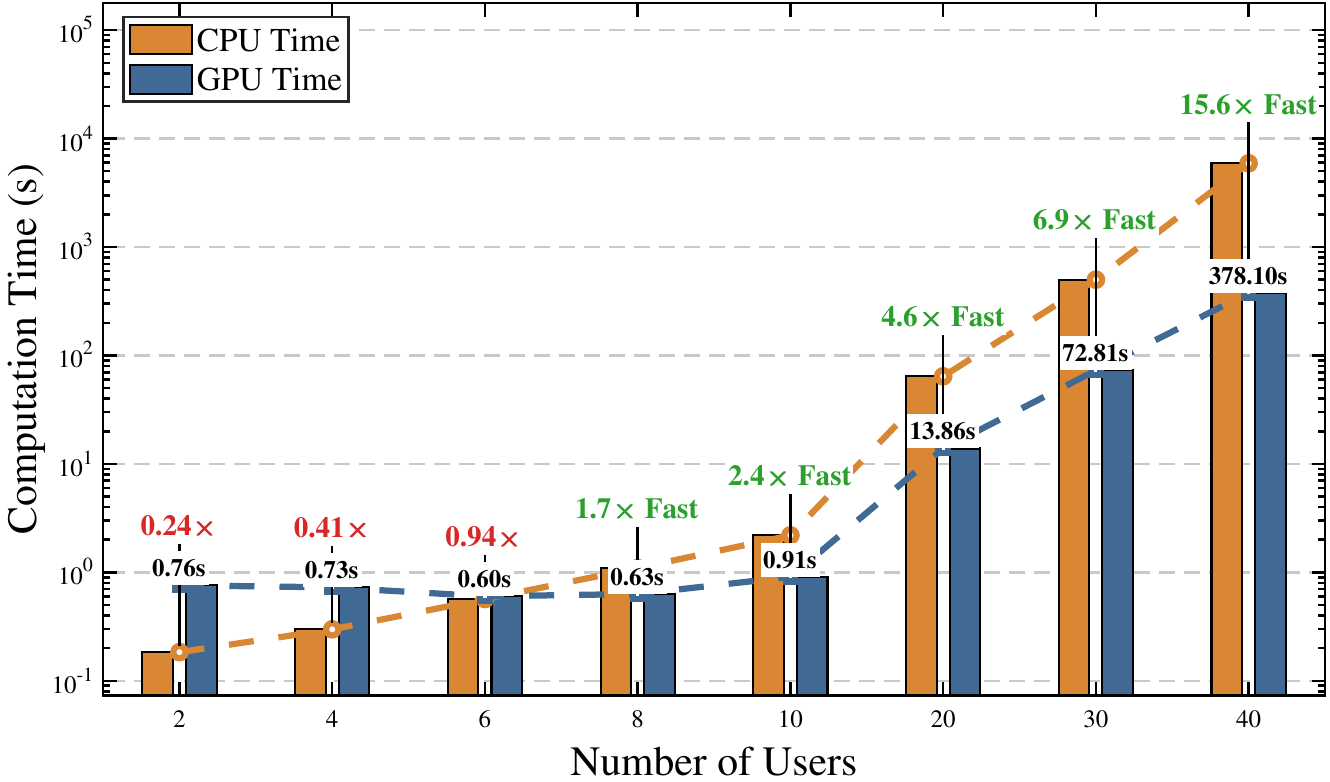}
    \caption{Execution time comparison between CPU and GPU versus the number of STs, assuming $M_t = 16$.}
    \label{fig:runtime_cpu_gpu_user}
\end{figure}
\begin{figure}[htbp]
    \centering
    \includegraphics[width=0.9\columnwidth]{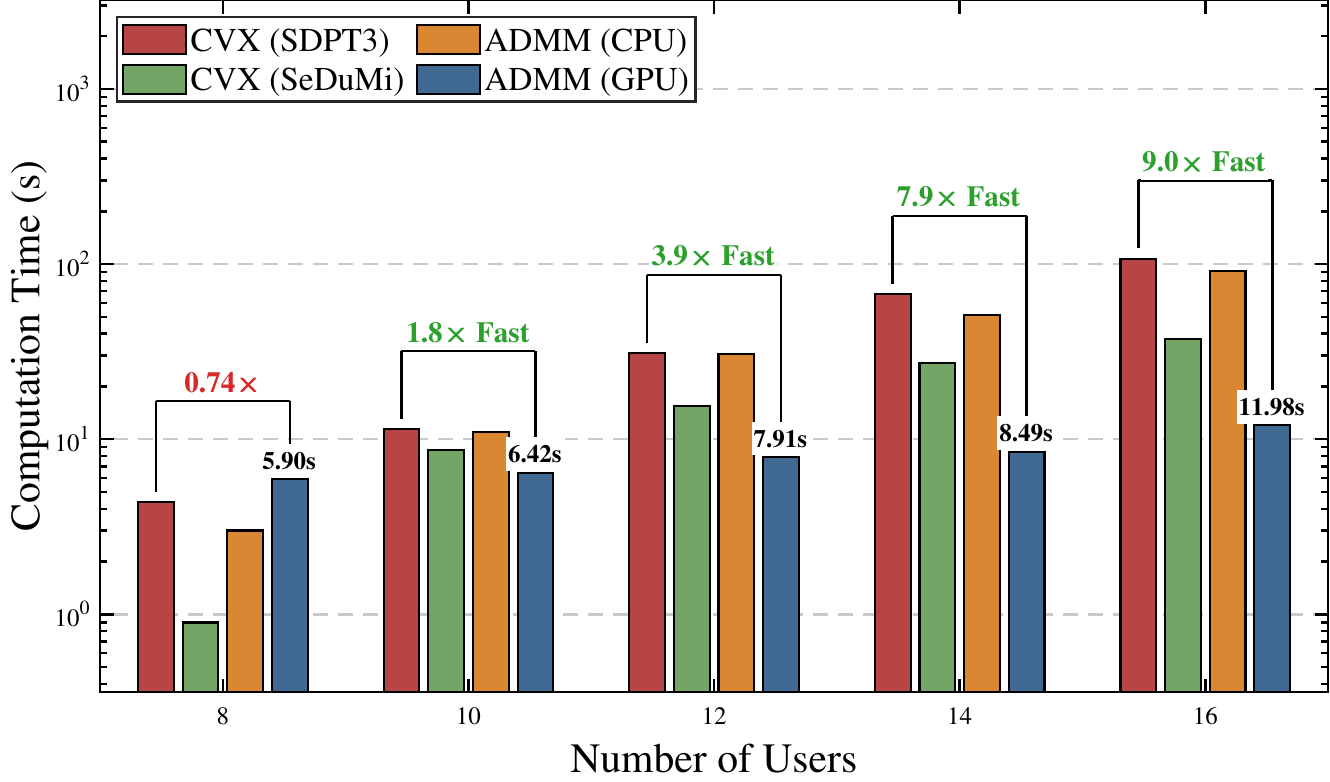}
    \caption{Execution time comparison between the proposed ADMM and the interior-point solvers versus the number of STs, assuming $M_t = 32$.}
    \label{fig:runtime_cpu_gpu_cvx_user}
\end{figure}

Finally, we evaluate the computational superiority of the proposed optimization framework. Fig. \ref{fig:runtime_cpu_gpu_user} compares the execution time between CPU and GPU versus the number of STs, assuming $M_t=16$. As the number of STs increases, the GPU implementation significantly outperforms the CPU, achieving a $15.6\times$ speedup when the system serves 40 STs. Fig. \ref{fig:runtime_cpu_gpu_cvx_user} extends the comparison to the conventional interior-point solvers under $M_t=32$. In the small-scale regime the GPU implementation still carries the kernel-launch and data-transfer overhead: with 8 STs it requires 5.90 s, which corresponds to $0.74\times$ the time of CVX-SDPT3, and the CPU implementation of ADMM is faster as well, since the variable dimension is too small to amortize the GPU overhead. This trend reverses as the problem grows. The GPU time increases only from 6.42 s to 11.98 s when the number of STs rises from 10 to 16, whereas CVX-SDPT3 increases from 11.38 s to 107.35 s and the CPU implementation from 11.03 s to 90.71 s over the same range. The resulting speedup over CVX-SDPT3 grows monotonically from $1.8\times$ at 10 STs to $3.9\times$, $7.9\times$ and $9.0\times$ at 12, 14 and 16 STs, respectively. Therefore, with the increase in problem dimensionality, GPU acceleration becomes essential when scaling up the array and the number of users.

\section{\uppercase{{\large C}onclusions}}
This paper investigated robust downlink beamforming for OTFS-enabled massive MIMO systems under 3D DDA-domain CSI uncertainty. We formulated a chance-constrained total transmit power minimization problem, decomposed the quadratic SINR terms by a Taylor expansion around the nominal channel while retaining a conservative bound on the residual term, and constructed SVC uncertainty sets from empirical CSI samples. Robust counterpart was derived for the model, and a GPU-accelerated ADMM algorithm was designed to solve the resulting convex programs efficiently. Simulation results demonstrated that the SVC uncertainty model provided the tightest robust feasible region, achieving up to 50\% power savings over the Box model while holding the achieved reliability at the prescribed level $1-\epsilon$. The results further showed that the GPU-accelerated ADMM scales far more gently than the conventional alternatives, achieving a $15.6\times$ speedup over the CPU implementation of the same algorithm with 40 STs and $M_t = 16$, and a $9.0\times$ speedup over the CVX/SDPT3 interior-point solver with 16 STs and $M_t = 32$; the advantage grows with the antenna array and the user population, at the price of a small overhead in the small-scale regime.

{\small
\bibliographystyle{IEEEtran}
\bibliography{reference}
}

\end{document}